\documentclass[twocolumn,showpacs,floatfix,eqsecnum,prb]{revtex4}

\usepackage[dvips]{epsfig}
\usepackage{amssymb}

\usepackage{float}

\begin{document}

\title{Critical Behavior of a Point Contact in a Quantum Spin Hall Insulator}

\author{Jeffrey C.Y. Teo and C.L. Kane}
\affiliation{Dept. of Physics and Astronomy, University of
Pennsylvania, Philadelphia, PA 19104}

\begin{abstract}
We study a quantum point contact in a quantum spin Hall insulator.
It has recently been shown that the Luttinger
liquid theory of such a structure maps to the theory of a weak link
in a Luttinger liquid with spin with Luttinger liquid parameters
$g_\rho = 1/g_\sigma = g < 1$.  We show that for weak interactions,
$1/2<g<1$, the pinch-off of the point contact as a function of gate
voltage is controlled by a novel quantum critical point, which
is a realization of a nontrivial intermediate fixed point found
previously in the Luttinger liquid model with spin.  We predict that the dependence of
the conductance on gate voltage near the pinch-off
transition for different temperatures collapses onto a universal curve
described by a crossover scaling function associated with that fixed point.
We compute the conductance and critical exponents of the critical
point as well as the universal scaling function in solvable limits,
which include $g=1-\epsilon$, $g=1/2+\epsilon$ and $g=1/\sqrt{3}$.
These results, along with a general scaling analysis provide an
overall picture of the critical behavior as a function of $g$.
In addition, we analyze the structure of the four terminal
conductance of the point contact in the weak tunneling and weak
backscattering limits.  We find that different components of the
conductance can have different temperature dependence.  In
particular, we identify a skew conductance $G_{XY}$, which we predict
vanishes as $T^\gamma$ with $\gamma\ge 2$.  This behavior is a direct
consequence of the unique edge state structure of the quantum spin
Hall insulator.  Finally, we show that
for strong interactions $g<1/2$ the presence of spin non conserving
spin orbit interactions leads to a novel time reversal symmetry
breaking insulating phase.  In this phase, the transport is carried
by spinless chargons and chargeless spinons.  These lead to
nontrivial correlations in the low frequency shot noise.
Implications for experiments on HgCdTe quantum well structures will be
discussed.

\end{abstract}

\maketitle

\section{Introduction}

A quantum spin Hall insulator (QSHI) is a time reversal invariant
two dimensional electronic phase which has
a bulk energy gap generated by the spin orbit interaction\cite{km1,bz}.  It has
a topological order\cite{km2} which requires
the presence of gapless edge states similar to those that occur in the
integer quantum Hall effect.  In the simplest version, the QSHI can be
understood as two time reversed copies of the integer quantum Hall state\cite{haldane} for up and down
spins.   The edge states, which propagate in opposite directions for the two
spins, form a unique one dimensional system in which elastic backscattering is
forbidden by time reversal symmetry\cite{km1}.  This state occurs in HgCdTe quantum well
structures\cite{bhz}, and experiments have verified the basic features of the edge
states, including the Landauer conductance\cite{molenkamp1} $2e^2/h$, as well as the non
locality of the edge state transport\cite{molenkamp2}.

In the presence of electron interactions, the edge states form a Luttinger
liquid\cite{zhangll,moorell,chamon1,strom,oreg,nagaosa3,ktlaw}.
For strong interactions (when the Luttinger liquid parameter $g < 3/8$)
random two particle backscattering processes destabilize the edge states,
leading to an Anderson localized phase.  For $g>3/8$ (or a sufficiently clean
system), however, one expects the characteristic power law behavior for
tunneling of a Luttinger liquid.

\begin{figure}
\centerline{ \epsfig{figure=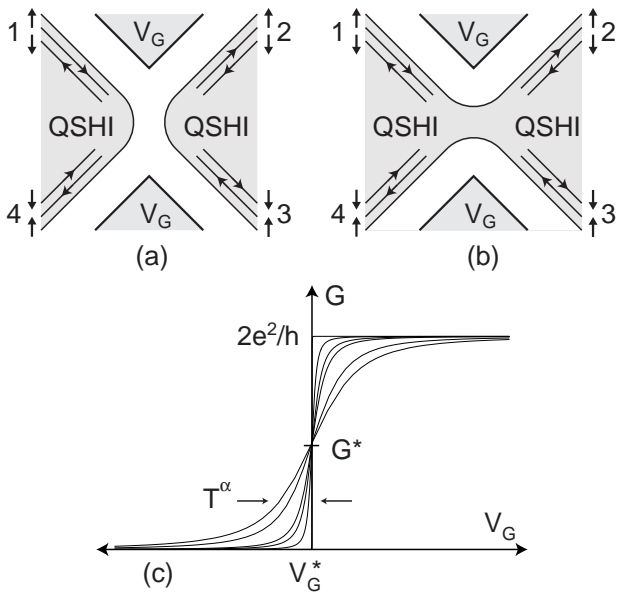,width=3.2in} }
 \caption{A quantum point contact in a QSHI, controlled by a gate
 voltage $V_G$.  In (a) $V_G<V_G^*$, and
 the point contact is pinched off.  The spin filtered
 edge states are perfectly reflected.
 In (b) $V_G>V_G^*$, and the point contact is open.  The edge states are
 perfectly transmitted.  In (c) we plot the conductance (later defined as $G_{XX}$) as
 a function of $V_G$ for different temperatures.  As the temperature is lowered, the
 pinch-off curve sharpens up with a width $T^\alpha$.  The curves cross at a
 critical conductance $G^*$, and the shape of the curve has the universal
 scaling form (\ref{scalingform}).   The plotted curves are based on Eq. \ref{g1-epsilon},
 valid for $g = 1-\epsilon$, which is computed in Section III.C. }
 \label{fig1}
\end{figure}

A powerful tool for probing edge state transport experimentally is to make a
quantum point contact.  As depicted in Fig. \ref{fig1}(a,b),
a gate voltage controls the coupling between edge
states on either side of a Hall bar as the point contact is pinched off.
Recently, the point contact problem for a QSHI
has been studied\cite{chamon1,strom}.
Hou, Kim and Chamon\cite{chamon1} made the interesting observation
that the QSHI problem maps
to an earlier studied model\cite{furusaki,kf1} of a weak link in
a spinful Luttinger liquid (SLL), in which the charge and spin Luttinger parameters
are given by $g_\rho = g$ and $g_\sigma = 1/g$ \cite{grhonote}.
For sufficiently strong interactions ($g<1/2$) they found that the
simple perfectly transmitting and perfectly reflecting phases are both
unstable.  They showed that as long as spin is conserved at the junction
the low energy behavior is dominated by a non trivial ``mixed" fixed point
of the SLL,
in which charge is reflected but spin is perfectly
transmitted.  This charge insulator/spin conductor (IC) phase leads to a
novel structure in the four terminal conductance of the point contact.

In this paper, we will focus on the QSHI point contact for weaker interactions, when
$1/2< g< 1$.  In this regime the open limit (or weak backscattering, ``small $v$") and the
pinched off limit (or weak tunneling, ``small $t$") are {\it both} stable perturbatively.  This
is different from the behavior in an ordinary Luttinger liquid\cite{furusaki,kf1,kf2} or a
fractional quantum Hall point contact\cite{moon,webb}.  In those cases the perfectly
transmitting limit is unstable for $g<1$.  Weak backscattering is relevant and
grows at low energy, leading to a crossover to the stable perfectly
reflecting fixed point.  The fact that both the small $v$ and the small $t$
limits are stable for the QSHI point contact
means that there must be an intermediate unstable fixed point
which separates the flows to the two limits.  This unstable fixed point
describes a quantum critical point where the point contact switches on as a
function of the pinch-off gate voltage.  We will argue that in the limit of
zero temperature the point contact switches on abruptly as a function of
gate voltage $V_G$, with
conductance $G=0$ for $V_G < V_G^*$ and $G= 2 e^2/h$ for $V_G > V_G^*$.  At
finite but low temperature $T$, the shape of the pinch-off curve $G(V_G,T)$
is controlled by the crossover between the unstable and stable fixed
points, and is described by a universal crossover scaling function,
\begin{equation}
\label{scalingform}
\lim_{\Delta V_G,T\rightarrow 0} G(V_G,T) =
{2 e^2\over h}{\cal G}_g(c{{\Delta V_G}\over T^{\alpha_g}}).
\end{equation}
Here $\Delta V_G = V_G - V_G^*$ and $c$ is a non universal constant.
$\alpha_g$ is a critical exponent describing the unstable intermediate fixed
point.  ${\cal G}_g(X)$ is a universal function which crosses over between $0$ and $1$
as a function of $X$.  $\alpha_g$ and ${\cal G}_g(X)$ are completely determined by
the Luttinger liquid parameter $g$.  This behavior means that as temperature is
lowered, the pinch-off curve as a function of $V_G$ sharpens up with a characteristic
width which vanishes as $T^{\alpha_g}$, as shown schematically in Fig. \ref{fig1}(c).  The curves at different
low temperatures cross at $G_g^* = {\cal G}_g(0)$, the conductance of the critical point.
Eq. \ref{scalingform} predicts that data from
different temperatures can be rescaled to lie on the same universal curve.

The
crossover scaling function ${\cal G}_g(X)$ is similar to the scaling
function that controls the
lineshape of resonances in a Luttinger liquid\cite{kf1,kf3} and in a fractional quantum Hall
point contact\cite{moon}.  That scaling function was computed exactly for all $g$ by
Fendley, Ludwig and Saleur\cite{fendley1} using the thermodynamic Bethe ansatz.  That problem,
however, was simpler than ours because the critical point occurs at the
weak backscattering limit, which is described by a boundary conformal
field theory with a trivial boundary condition\cite{boundarycft}.
The intermediate fixed point relevant to our problem has no such
simple description.  Thus, even the critical point properties $\alpha_g$ and
$G_g^*$ (which were simple for the resonance problem) are highly nontrivial to
determine.

Intermediate fixed points in Luttinger liquid problems
were first discussed in Refs. \onlinecite{furusaki} and \onlinecite{kf1} in the context
of SLLs.
However, for that problem they occur in a rather unphysical region of parameter
space $g_\sigma > 2$, because spin rotational invariance requires $g_\sigma = 1$.
To our knowledge, the QSHI point contact provides the first
physically viable system to directly probe these non trivial fixed points.

The existence of the intermediate fixed points
can be inferred from the stability of the simple perfectly transmitting or
reflecting fixed points\cite{furusaki,kf1}.
However their properties are difficult to compute, and a general
characterization of these critical points remains an unsolved problem in
conformal field theory\cite{affleck1}.  Two approaches have been used to study their
properties.   In Ref. \onlinecite{kf1}, a perturbative
approach was introduced which applies when the Luttinger parameters
are close to their critical values $g_\sigma^*$ and $g_\rho^*$, where the simple fixed points
become unstable.  (For instance, for the weak backscattering limit, $g_\rho^* = 1/2$, $g_\sigma^* = 3/2$).
For $g_{\rho,\sigma} = g_{\rho,\sigma}^*-\epsilon$, the fixed point is accessible in perturbation
theory about the simple fixed point, and it's properties can be computed in a
manner analogous to the $\epsilon$ expansion in statistical mechanics.

An alternative approach is to map the theory for specific values of $g_\rho$
and $g_\sigma$ onto solvable models.  In Ref. \onlinecite{yikane},
Yi and Kane recast the Luttinger liquid barrier
problem as a problem of quantum Brownian motion (QBM) in a
two dimensional periodic potential.  When $g_\rho=1/3$, $g_\sigma=1$ and
the potential has minima with a honeycomb lattice
symmetry, a stable intermediate fixed point which occurs in that problem
was identified with that of the 3
channel Kondo problem.  This, in turn is related to the solvable $SU(2)_3$ Wess Zumino
Witten model\cite{ludwig}, allowing for a complete characterization of the fixed point.
This idea was further developed by Affleck, Oshikawa and Saleur\cite{affleck1}, who provided a more
general characterization of the fixed point in terms of the boundary conformal field
theory of the three state Pott's model.
For $g_\rho = 1/\sqrt{3}$ and $g_\sigma = \sqrt{3}$ the
QBM model with triangular lattice symmetry has
an unstable intermediate fixed point, which we will
see is related to the fixed point of the QSHI problem.  In Ref. \onlinecite{yikane} symmetry arguments were
exploited to determine the critical conductance $G^*$ in that case.

In this paper we will compute $\alpha_g$ and ${\cal G}_g(X)$ (along with a
multiterminal generalization of the conductance) in three solvable limits:

{\it (i)} For
$g=1-\epsilon$, we will perform an expansion for weak electron interactions.  For non
interacting electrons the point contact can be characterized in
terms of a scattering matrix $S_{ij}$, for arbitrary transmission.
Weak interactions lead to a logarithmic
renormalization of $S_{ij}$.  Following the method developed by Matveev,Yue
and Glazman\cite{glazman}, this allows ${\cal G}_g(X)$ and $\alpha_g$ to be calculated exactly in the limit
$g\rightarrow 1$.

{\it (ii)} For $g = 1/2+\epsilon$ we find that the intermediate fixed point
approaches the charge insulator/spin conductor fixed point, allowing for a
perturbative calculation of the fixed point properties $G_g^*$ and $\alpha_g$ to leading order
in $\epsilon$.  Moreover, for $g=1/2$
the Luttinger liquid theory can be fermionized, which allows the full
crossover function ${\cal G}_g(X)$ to be determined in that limit.

{\it (iii)} For $g = 1/\sqrt{3}$ the self duality argument developed in Ref.
\onlinecite{yikane}
allows us to compute the fixed point conductance $G^*$ exactly.

These three results, along with the general scaling analysis provide an overall
picture of the critical behavior of the QSHI point contact as a function of
$g$.

In addition to the analysis of the pinch-off transition discussed above, we
will touch on two other issues in this paper.  First, we will introduce a convenient
parameterization of the four terminal conductance as a $3 \times 3$ conductance
matrix.  In this form symmetry
constraints on the conductance are reflected in a natural way.  Moreover, we will predict that
different components of the conductance matrix have different temperature
dependence at the low temperature fixed points.  In particular, we will introduce
a ``skew" conductance $G_{XY}$,
which is predicted to vanish as $T^{\gamma}$ with $\gamma \ge 2$.
For non interacting electrons we will show that $G_{XY}=0$, and for
weak interactions $\gamma = 2$.
This behavior is a direct consequence of the spin filtered nature of the
edge states, and does not occur in a generic four terminal conductance device.
It is thus a powerful diagnostic for the edge states.

Secondly, we will examine the role of spin orbit terms
at the point contact which respect time reversal symmetry but violate spin conservation.
For $g>1/2$ we will provide evidence that such terms are
{\it irrelevant} at the intermediate critical fixed point, so that they
are unimportant for the critical behavior of the point
contact.  However, for $g<1/2$, such terms are relevant.  Hou, Kim and
Chamon\cite{chamon1} pointed out that these terms are relevant perturbations at the charge
insulator/spin conductor fixed point for $g<1/2$, but they did not identify the
stable phase to which the system flows at low energy.  We will argue that the
system flows to a time reversal symmetry breaking insulating state in which the {\it four terminal}
conductance $G_{ij}=0$.  Since spin orbit interaction terms will generically be
present in a point contact, the true low energy behavior of a point contact
will be described by this phase.  An interesting consequence of the broken time
reversal symmetry of this phase is that the weak tunneling processes which
dominate the conductance at low, but finite temperature are not electron
tunneling processes.  Rather, they involve the tunneling of neutral spinons and
spinless chargons.  This has nontrivial implications for four terminal noise
correlation measurements.  A related effect has been predicted by
Maciejko et al.\cite{oreg} for the insulating state of a single
impurity on a single edge of a QSHI.  This insulating state, however,
requires stronger electron electron interactions.  It occurs in the
regime $g<1/4$, where weak disorder already leads to Anderson
localization.

This paper is organized as follows.  In section II we discuss our model and
analyze five stable phases.  In addition to the simple fixed points, where
charge and spin are either perfectly reflected or perfectly transmitted, we
discuss the time reversal symmetry breaking insulating phase which
occurs for strong interactions with spin orbit.  In section III we discuss the
critical behavior of the conductance at the pinch-off transition.
We will begin in section III.A with a
general discussion of the scaling theory and phase diagram,
along with a summary of our results.  Readers who are not interested in the
detailed calculations can go directly to this subsection.
In the following subsections we describe
the calculations for $g=1/\sqrt{3}$, $g=1-\epsilon$ and $g=1/2+\epsilon$ in
detail.  In section IV we conclude with a discussion of experimental and
theoretical issues raised by this work.  In appendix A we show describe our
parameterization of the four terminal conductance and show that in this representation
symmetry constraints have a simple form.

\section{Model and Stable Phases}

In this section we will describe the Luttinger liquid theory of the QSHI point
contact.  We will begin in section II.A by describing the Luttinger
liquid model first for a single edge and then relating the four edges
to the theory of the SLL.  We then discuss the
four terminal conductance.  In section II.B we describe the simple
limits of our model which correspond to stable
phases.  The simplest limits are the perfect transmission limit, or charge
conductor/spin conductor (CC), the perfect reflection limit, or
charge insulator/spin insulator (II).  In addition we will discuss
the ``mixed" phases, including the charge insulator/spin conductor
(IC) and the charge conductor/spin insulator (CI).

For most of this section we will assume that spin is conserved.
While spin nonconserving spin orbit interactions are allowed and will
generically be present we will argue that they are {\it irrelevant} for the
fixed points and crossovers of physical interest.  An exception to
this, however, occurs for strong interactions when $g<1/2$.  This
 will be discussed in section II.B.5, where
 we will show that there are relevant spin orbit terms which
destabilize the CC, II and IC phases.  We will argue that these
perturbations flow to a different low temperature phase, which we
identify as a time reversal symmetry breaking insulator (TBI).  In
that section we will explore the transport properties of that state.

Much of the theory presented in this section is contained either explicitly or
implicitly in the work Hou, Kim and Chamon\cite{chamon1},
as well as in Refs. \onlinecite{strom,furusaki,kf1}.  We include it
here to establish our notation and to make our discussion self contained.
We will highlight, however, three results of this section which are original to this work.
They include (1)  our analysis of the four terminal conductance, which predicts
that different components of the conductance matrix have different temperature
dependence.  In particular, we find that the skew conductance $G_{XY}$ vanishes
at low temperature as $T^{\gamma}$ with $\gamma\ge 2$.
(2) In section II.B.5 we introduce the TBI phase discussed above.
(3) We introduce a perturbative analysis of the IC and
CI phases in section II.B.3 and II.B.4.  While this was
partially discussed in Ref. \onlinecite{kf1}, we will show that a full analysis
requires the introduction of a pseudo-spin degree of freedom in the
perturbation theory.  This new pseudo-spin does not affect the lowest
order stability analysis of the IC phase, but it will prove crucial
for the second order renormalization group flows, which will be used
in the $\epsilon$ expansion in Section III.D.

\subsection{Model}

The edge states on the four edges in Fig. \ref{fig1}(a,b)
emanating from the point contact may be
described by the Hamiltonian
\begin{equation}
H_0 = \sum_{i=1}^4 \int_0^\infty dx_i {\cal H}_0^i,
\end{equation}
with
\begin{eqnarray}
{\cal H}_0^i &=&  iv_0(\psi_{i,\rm
in}^\dagger\partial_x\psi_{i,\rm in} - \psi_{i,\rm
out}^\dagger\partial_x\psi_{i,\rm out}) \nonumber \\
 &+& u_2 \psi_{i,\rm in}^\dagger \psi_{i,\rm in}
\psi_{i,\rm out}^\dagger \psi_{i,\rm out}  \\
&+& {1\over 2} u_4 \left[( \psi_{i,\rm in}^\dagger \psi_{i,\rm in})^2 +
(\psi_{i,\rm out}^\dagger \psi_{i,\rm out})^2\right]. \nonumber
\label{h0psi}
\end{eqnarray}
Here $\psi_{i,\rm in}$ and $\psi_{i,\rm out}$ are a time reversed pair of
fermion operators with opposite spin which propagate toward and away from the
junction.  $v_0$ is the bare Fermi velocity, and
$u$ is electron interaction strength.  $u_2$ and $u_4$ are forward scattering
interaction parameters.  The boundary condition on the
fermions at $x=0$ is determined by the transmission of the point contact, and
will be discussed in various limits below.

\subsubsection{Bosonization of a single edge}

We first consider the Luttinger liquid theory for a single edge.  We thus
bosonize according to
\begin{equation}
\label{bosonizepsi}
\psi_{i, a} = {1\over\sqrt{2\pi x_c}} e^{i\phi_{i, a}},
\end{equation}
where $a = {\rm in}, {\rm out}$, and $x_c$ is a short distance cutoff.
$\psi_{i,a}$ obey the Kac Moody commutation algebra,
\begin{equation}
\label{kacmoodyphi}
[\phi_{i,a}(x), \phi_{j,b}(y)] = i \pi  \delta_{ij}\tau^z_{ab}{\rm sgn}(x-y).
\end{equation}
Then,
\begin{eqnarray}
\label{h0phi}
{\cal H}_0^i &=& {v_0\over{4\pi}}
\left[ (1+\lambda_4)\left((\partial_x\phi_{i, \rm in})^2 +
(\partial_x\phi_{i, \rm out})^2\right)\right. \nonumber\\
&&- \left. 2\lambda_2 \partial_x\phi_{i,\rm in} \partial_x\phi_{i,\rm
out}\right],
\end{eqnarray}
where $\lambda_i = u_i/(2\pi v_0)$.
Changing variables
\begin{equation}
\left(\begin{array}{l} \phi_{i,\rm in} \\ \phi_{i, \rm out} \end{array}\right)
=
{1\over {2g}} \left(\begin{array}{ll} 1+g & 1-g \\
 1-g & 1+g \end{array}\right)
 \left(\begin{array}{l} \tilde\phi_{i,\rm in} \\  \tilde\phi_{i, \rm
 out}\end{array}\right)
\end{equation}
transforms (\ref{h0phi}) into a theory of
decoupled chiral bosons
\begin{equation}
\label{h0phitilde}
{\cal H}_0 = {v\over{4\pi g}}\left[(\partial_x\tilde\phi_{i, \rm in})^2 +
(\partial_x\tilde\phi_{i, \rm out})^2 \right],
\end{equation}
where $\tilde\phi_{i,a}$ obey
\begin{equation}
\label{kacmoodyphitilde}
[\tilde\phi_{i,a}(x), \tilde\phi_{j,b}(y)] = i \pi g \delta_{ij}\tau^z_{ab}{\rm sgn}(x-y).
\end{equation}
Here $v = v_0\sqrt{(1+\lambda_4)^2-\lambda_2^2}$ and
\begin{equation}
\label{g}
g = \sqrt{1+\lambda_4- \lambda_2 \over{1+\lambda_4+\lambda_2}}.
\end{equation}
The Luttinger liquid parameter $g$ determines the power law exponents for various quantities.  For
instance the tunneling density of states scales as $\rho(E) \propto
E^{(g+1/g)/2-1}$.

\subsubsection{Mapping to Spinful Luttinger liquid}

Consider an open point contact in a Hall bar geometry with edge states on the top and bottom
edges which continuously connect leads 1 and 2 and leads 3 and 4.  We then define left and
right moving fields with spin $\uparrow,\downarrow$ as
\begin{eqnarray}
\label{phirl}
\phi_{R\uparrow} &= \phi_{1,\rm in}(-x) \theta(-x) + \phi_{2,\rm out}(x) \theta(x) \nonumber\\
\phi_{L\downarrow} &= \phi_{2,\rm in}(x) \theta(x) +\phi_{1,\rm out}(-x) \theta(-x) \nonumber  \\
\phi_{L\uparrow} &=  \phi_{3,\rm in}(x) \theta(x)+\phi_{4,\rm out}(-x) \theta(-x)  \\
\phi_{R\downarrow} &= \phi_{4,\rm in}(-x) \theta(-x) + \phi_{3,\rm out}(x)
\theta(x).\nonumber
\end{eqnarray}
It is then useful to define sum and difference fields as
\begin{equation}
\label{phiasigma}
\phi_{a\sigma} = {1\over 2}(\varphi_\rho + \sigma \varphi_\sigma + a\theta_\rho
+ a\sigma \theta_\sigma),
\end{equation}
where $a=R,L= +,-$ and $\sigma=\uparrow,\downarrow=+,-$.
Then, $\theta_\alpha$ and $\varphi_\alpha$
obey,
\begin{equation}
[\theta_\alpha(x),\varphi_\beta(y)] = 2\pi i
\delta_{\alpha\beta}\theta(x-y),
\end{equation}
and (\ref{h0psi}, \ref{h0phi}) become\cite{chamon1}
\begin{equation}
\label{h0thetaphi}
H_0 = \int_{-\infty}^\infty dx \sum_{a=\sigma,\rho} {v\over{4\pi}}\left[ g_a (\partial_x\varphi_a)^2
+ {1\over g_a} (\partial_x\theta_a)^2 \right].
\end{equation}
where
\begin{equation}
\label{grhogsigma}
g_\rho = g, \quad g_\sigma = 1/g,
\end{equation}
and $g$ and $v$ are given in the previous section.

It is useful to list the effect of symmetry operations on the
charge-spin variables, because symmetries constrain the allowed
tunneling operators.  Charge conservation leads to gauge invariance
under  the transformation $\varphi_\rho \rightarrow \varphi_\rho +
\delta_\rho$.  The conservation of spin $S_z$ leads to invariance under
$\varphi_\sigma \rightarrow \varphi_\sigma + \delta_\sigma$.  The effects
of time reversal and mirror symmetries is shown in Table I.
Time reversal symmetry is specified by the operation
$\Theta \psi_{a\sigma}\Theta^{-1} = i \sigma \psi_{\bar
a\bar\sigma}$.  The mirror ${\cal M}_X$
interchanges leads $14 \leftrightarrow 23$ while
${\cal M}_Y$ interchanges leads $(12 \leftrightarrow 34)$.

\begin{table}
\centering
\begin{tabular}{|c|c|c|c|}\hline
$O$ & $\Theta O \Theta^{-1}$ & ${\cal M}_X O {\cal M}_X^{-1}$ & $ {\cal M}_Y O {\cal M}_Y^{-1}$\\
\hline
$\theta_\rho$ & $\theta_\rho$ & $-\theta_\rho$ & $\theta_\rho$ \\
$\varphi_\rho$ &$ - \varphi_\rho$ & $\varphi_\rho$ & $\varphi_\rho$ \\
$\theta_\sigma$ & $- \theta_\sigma$ & $\theta_\sigma$ & $- \theta_\sigma$ \\
$\varphi_\sigma$ & $\varphi_\sigma+\pi$ & $-\varphi_\sigma$ & $- \varphi_\sigma$ \\
  \hline
\end{tabular}
  \caption{The effect of discrete symmetry operations on the boson fields $\theta_\rho$ and
  $\theta_\sigma$.}
  \label{symtab}
\end{table}

%

\subsubsection{Four Terminal Conductance}

The central measurable quantity is the four terminal conductance, defined by
\begin{equation}
I_i = \sum_j G_{ij} V_j,
\end{equation}
where $I_i$ is the current flowing into lead $i$.
$G_{ij}$ is in general characterize by 9 independent parameters.  In Appendix
A we introduce a convenient representation for these parameters,
which simplifies the representation of symmetry constraints.  Here we will
summarize the key points of that analysis.

The presence of both time reversal symmetry and spin conservation
considerably simplifies the conductance.  It is characterized by {\it
three} independent conductances
\begin{equation}
\label{ixiy}
\left(\begin{array}{c} I_X \\ I_Y \end{array}\right) =
\left(\begin{array}{cc} G_{XX} & G_{XY} \\  G_{YX} & G_{YY} \end{array}\right)
\left(\begin{array}{c} V_X \\ V_Y \end{array}\right).
\end{equation}
Here $I_X = I_1+I_4$ is the current flowing from left to right in Fig. 1,
while $I_Y = I_1 + I_2$ is the current flowing from top to bottom.
Similarly, $V_X$ is a voltage biasing leads (14) relative to (23) and
$V_Y$ biases leads (12) relative to (34).  $G_{XX}$ is thus the {\it
two terminal} conductance measured horizontally, while $G_{YY}$ is
the two terminal conductance measured vertically.  $G_{XY}=G_{YX}$ is
a ``skew conductance", which vanishes in the presence of mirror
symmetry.  Given these three parameters, the full four terminal
conductance matrix $G_{ij}$ can be constructed using Eq. (\ref{gijgab}).

A second consequence of spin conservation is the quantization of a
particular combination of $G_{ij}$.  In particular, in appendix A we
define a third current $I_Z = I_1 + I_3$ and a third voltage $V_Z$ which
biases leads (13) relative to (24).   Spin conservation then requires
\begin{equation}
\label{izgzzvz}
I_Z =G_{ZZ}  V_Z,
\end{equation}
with
\begin{equation}
\label{gzz}
G_{ZZ} =2 {e^2\over h}.
\end{equation}
Since spin nonconserving spin orbit terms are allowed, spin
conservation will not be generically present in the microscopic
Hamiltonian of the junction.  Nonetheless, we will argue that the low
temperature fixed points possess a {\it emergent} spin conservation,
as well as mirror symmetry, so that (\ref{gzz}) should hold, albeit with
corrections which vanish as a power of temperature.



\subsection{Stable Phases}

In this section we describe various stable fixed points which admit simple
descriptions using bosonization.  We will first focus on the limit in which
spin is conserved at the junction.  There are then four simple fixed points\cite{furusaki,kf1}.
These include the perfectly transmitting (CC)
limit, in which both charge and spin conduct, and the perfectly reflecting
limit (II) in which both charge and spin are insulating.  The ``mixed" fixed
points, denoted IC (CI) are perfectly reflecting for charge (spin) and
perfectly transmitting for spin (charge).

In the presence of spin non conserving spin orbit terms (which preserve time
reversal symmetry) an additional fixed point is possible in which time reversal
symmetry is spontaneously broken.  We will see that in the presence of
spin orbit terms this time reversal breaking insulator
(TBI) phase is the stable phase when $g<1/2$.

\subsubsection{Weak backscattering (CC) limit}



We first consider the limit where the point contact is nearly open and assume
spin is conserved.  It will prove
useful to follow Ref. \onlinecite{kf1} and write (\ref{h0thetaphi})
as a 0+1 dimensional Euclidean
path integral for $\theta_{\rho,\sigma}(\tau) \equiv
\theta_{\rho,\sigma}(x=0,\tau)$.  This formulation is not essential
for carrying out the perturbative analysis of this fixed point.  However, it is
of conceptual value for discussing the duality between different phases,
which can be understood in terms of instanton processes in which
$\theta_{\rho,\sigma}(\tau)$ tunnels between degenerate minima at strong coupling.
This is accomplished by setting up the path
integral for $\theta_{\sigma\rho}(x,\tau)$ and then integrating out
$\theta_{\sigma,\rho}(x,\tau)$ for $x\ne 0$.  The resulting theory for
$\theta_{\sigma,\rho}(\tau)$ has the form of a {\it quantum Brownian
motion} model\cite{schmid,fz,guinea,yikane,affleck1}, described by the Euclidean action
\begin{equation}
\label{scc}
S_{CC} = {1\over\beta}\sum_{\alpha,\omega_n}{1\over {2\pi g_\alpha}} |\omega_n|
|\theta_a(\omega_n)|^2 - \int_0^\beta {d\tau\over{\tau_c}}
V_{CC}(\theta_\sigma,\theta_\rho),
\end{equation}
where $\omega_n= 2\pi n/\beta$ are Matsubara frequencies, and $\beta = 1/k_BT$.
We have
included the short time cutoff $\tau_c = x_c/v$ in the second term to make the
potential $V(\theta_\rho,\theta_\sigma)$ dimensionless.  The theory
can be regularized by evaluating frequency sums with a
$\exp(-|\omega_n|\tau_c)$ convergence factor.

\begin{figure}
\centerline{ \epsfig{figure=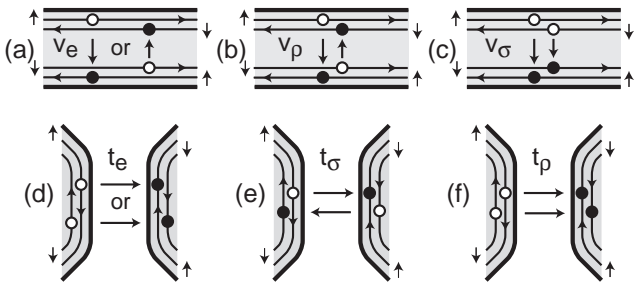,width=3.3in} }
 \caption{Schematic representation of tunneling processes in (a,b,c) the
 CC phase (``small $v$") and (c,d,e) the II phase (``small $t$").  (a) and (d)
 describe single electron processes, while the others are two particle processes.
 The duality relating $v_e \leftrightarrow t_e$, $v_\rho \leftrightarrow t_\sigma$ and
 $v_\sigma \leftrightarrow t_\rho$ can clearly be seen.}
 \label{tunnelingop}
\end{figure}

The potential $V(\theta_\rho,\theta_\sigma)$ is given by an expansion in terms
of tunneling operators, which represent the processes depicted in Fig.
\ref{tunnelingop}(a,b,c),
\begin{equation}
\label{vcc}
V_{CC} =  v_e \cos(\theta_\rho+\eta_\rho) \cos\theta_\sigma
+  v_\rho \cos 2\theta_\rho+ v_\sigma \cos 2\theta_\sigma.
\end{equation}
$v_e$ represents the elementary backscattering of a single electron
across the point contact.
The phase of $\cos\theta_\sigma$ in that term is fixed by time reversal
symmetry.  The phase $\eta_\rho$ of $\cos\theta_\rho$ is arbitrary, though mirror symmetry,
if present, requires $\eta_\rho = n\pi$.
In addition we include compound tunneling processes.
$v_\rho$ represents the backscattering of a pair
of electrons with opposite spins.  We have chosen to define $\theta_\rho$ such that the
phase of this term is zero.  Note that this process involves the
tunneling of {\it spin} (not charge) between the top and bottom edges.
Similarly, $v_\sigma$ represents the transfer of a unit of spin from the
right to the left moving channels, and involves the tunneling of {\it charge}
$2e$ between the top and bottom edges.  In general higher order terms could
also be included.  However, those terms are less relevant.

The low energy stability of this fixed point is determined by the scaling
dimensions $\Delta(v_\alpha)$ of the perturbations, which determine the leading
order renormalization group flows,
\begin{equation}
\label{dvdl}
dv_\alpha/d\ell = (1-\Delta(v_\alpha)) v_\alpha.
\end{equation}
These are given by
\begin{eqnarray}
\label{delta(v)}
\Delta(v_e) =& (g_\rho + g_\sigma)/2 &= (g+g^{-1})/2 \nonumber \\
\Delta(v_\rho) =& 2 g_\rho &= 2 g \\
\Delta(v_\sigma) =& 2 g_\sigma &= 2 g^{-1}.\nonumber
\end{eqnarray}
It is therefore clear that all operators are irrelevant for $1/2<g<2$, so that
the CC phase is stable.
For $g<1/2$ $v_\rho$ becomes relevant, and for $g>2$ $v_\sigma$ becomes
relevant.

At the fixed point the conductance matrix elements are
\begin{eqnarray}
\label{ccgxy}
G_{XX} &=& 2 e^2/h\nonumber \\
G_{YY} &=& G_{XY} = 0.
\end{eqnarray}
At finite temperature, there will be corrections to these values.  The leading
corrections will depend on the least irrelevant operators.  We find
\begin{eqnarray}
 \label{ccdeltag}
\delta G_{XX} &=& \left\{\begin{array}{ll}
 - c_1 v_e^2 T^{g+g^{-1} -2}
& g > 1/\sqrt{3} \\
- c_2 v_\rho^2 T^{4g -2} & g < 1/\sqrt{3}\end{array}\right. \nonumber \\
\delta G_{YY} &=& \left\{\begin{array}{ll}
c_3 v_e^2 T^{g+g^{-1}-2} & g < \sqrt{3} \\
 c_4 v_\sigma^2  T^{4/g-2} & g > \sqrt{3}\end{array}\right. ,
\end{eqnarray}
where $c_i$ are nonuniversal constants.  Note that for $g<1/\sqrt{3}$ the
exponents for $G_{XX}$ and $G_{YY}$ are different.  In addition,
there will be power law corrections to $G_{XY}$ when the mirror
symmetries ${\cal M}_x, {\cal M}_y$ are violated.  However, this correction
is zero when computed from (\ref{scc},\ref{vcc}),
even when $\eta_\rho \ne 0$, due to the symmetry of (\ref{vcc}) under
$\theta_\sigma \rightarrow - \theta_\sigma$.  Computing $G_{XY}$
requires a higher order irrelevant operator.  For instance
$\lambda_1 \partial_x\varphi_\sigma
\sin\theta_\rho \cos\theta_\sigma$ and
$\lambda_2 \partial_x\varphi_\rho \cos\theta_\rho \sin \theta_\sigma
$
break both ${\cal M}_X$ and ${\cal M}_Y$, while preserving
time reversal.  This leads to
\begin{equation}
\label{ccdeltagxy}
\delta G_{XY} = c_5 \lambda_1 \lambda_2 T^{g+g^{-1}}.
\end{equation}
Note that the temperature exponent of $G_{XY}$ is at least $2$ -
even for weak interactions $g\sim 1$.  This is because the tunneling terms
$\lambda_1$ and $\lambda_2$ include an extra derivative term.  This is related
to the fact (which we will show in Section III.C) that
for non interacting electrons $G_{XY} = 0$.  Weak interactions then
introduce inelastic processes which give $G_{XY} \propto T^2$.
The vanishing of $G_{XY}$ is a
unique property of the spin filtered edge states of the QSHI, which does not occur
for a generic four terminal conductance.

\subsubsection{Weak Tunneling (II) limit}

When the point contact is pinched off, $\theta_{\rho,\sigma}$ are effectively
pinned, and a theory can be developed in terms of electron tunneling process
across the point contact.  This theory is most conveniently expressed in terms
of the {\it discontinuity} $\tilde\theta_{\sigma,\rho} \equiv \varphi_{\sigma,\rho}^{\rm right} -
\varphi_{\sigma,\rho}^{\rm left}$ across the junction\cite{varphi}.
The theory takes the form
\begin{equation}
\label{sii}
S_{II} = {1\over\beta}\sum_{\alpha,\omega_n}{g_\alpha \over {2\pi }} |\omega_n|
|\tilde\theta_a(\omega_n)|^2 - \int_0^\beta
 {d\tau\over\tau_c} V_{II}(\tilde\theta_\sigma,\tilde\theta_\rho),
\end{equation}
with
\begin{equation}
\label{vii}
V_{II} =  t_e \cos(\tilde\theta_\rho+\eta_\rho) \cos\tilde\theta_\sigma
+  t_\rho \cos 2\tilde\theta_\rho+ t_\sigma \cos 2\tilde\theta_\sigma.
\end{equation}
As depicted in Fig. \ref{tunnelingop}(d,e,f)
$t_e$ represents the tunneling of a single electron from left to right across
the junction.  $t_\sigma$
describes the transfer of a unit of spin across the junction.  $t_\rho$
describes the tunneling of a pair of electrons with opposite spins.

The relationship between $S_{II}$ and $S_{CC}$ can be understood in two ways.
First, since both $S_{II}$ and $S_{CC}$ describe tunneling between the middles of two
disconnected Luttinger liquids (either on the top and bottom of the junction or
the left and right) the two theories are identical.  It is straightforward to
see that if we make the identification
\begin{eqnarray}
\label{thetadual}
\theta_\rho & \leftrightarrow \tilde\theta_\sigma \nonumber\\
\theta_\sigma & \leftrightarrow \tilde\theta_\rho
\end{eqnarray}
it follows that
\begin{equation}
\label{siiccdual}
S_{II}(g_\rho,g_\sigma,t_e,t_\rho,t_\sigma) =
S_{CC}(g_\sigma,g_\rho,v_e,v_\sigma,v_\rho).
\end{equation}
Thus, the ``small v" and ``small t" theories are dual to each other, with the
identification
\begin{eqnarray}
\label{vtdual}
v_e &\leftrightarrow& t_e\nonumber\\
v_\rho &\leftrightarrow& t_\sigma\nonumber\\
v_\sigma &\leftrightarrow& t_\rho \\
g &\leftrightarrow& g^{-1}.\nonumber
\end{eqnarray}
Using this identification, the scaling dimensions $\Delta(t_\alpha)$ can be
read off from Eq. \ref{delta(v)}.  Thus, like the CC phase, the II phase is stable when
$1/2<g<2$.  The low temperature conductance can also be read from (\ref{ccgxy}),
(\ref{ccdeltag}) and (\ref{ccdeltagxy}) using the identification
\begin{equation}
G_{XX} \leftrightarrow G_{YY}.
\end{equation}

\begin{figure}
\centerline{ \epsfig{figure=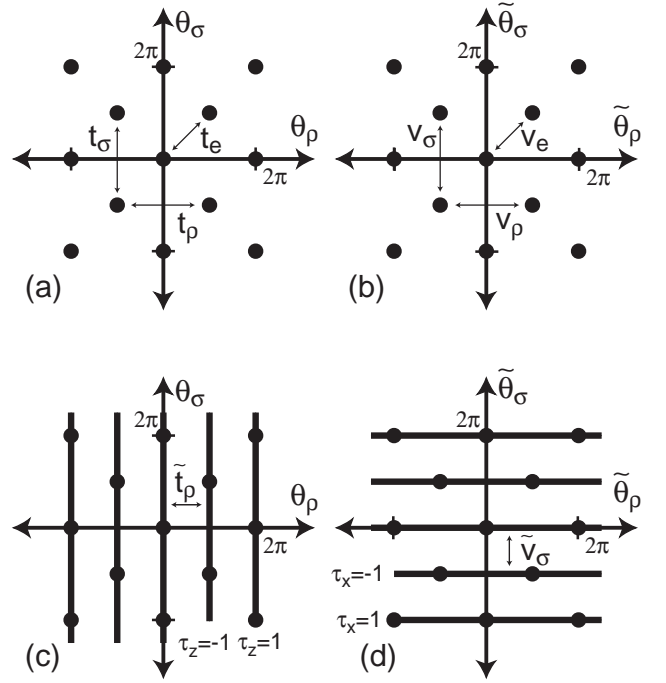,width=3.3in} }
 \caption{(a) Positions of the minima of $V(\theta_\rho,\theta_\sigma)$ in Eq.
 \ref{vcc}.  When the minima are deep instanton tunneling events
  between the minima, denoted by $t_e$,
 $t_\rho$ and $t_\sigma$ correspond to the transfer of charge and spin across
 the junction, and define the dual theory (\ref{sii},\ref{vii}).  (b)  Positions of the minima
 of $V(\tilde\theta_\rho,\tilde\theta_\sigma)$ in the dual theory (\ref{sii},\ref{vii}).
 Instanton process $v_e$, $v_\rho$ and $v_\sigma$ correspond to backscattering of
 charge and spin in the original theory.  (c) The IC phase viewed from the
 CC limit.  When $v_\rho$ is large and $v_\sigma=0$,
 the minima of $V(\theta_\rho,\theta_\sigma)$ in Eq.
 \ref{vcc} are on one dimensional valleys, and define the IC phase.  When $v_\sigma$ is
 small but finite the valleys have a periodic potential
 $\tilde v_\sigma\tau^z \cos\theta_\sigma$,
 with opposite signs $\tau^z=\pm 1$ in neighboring valleys.  Instanton tunneling processes
 between the valleys, denoted $\tilde t_\rho$, switch the sign of $\tau^z$.
 (d)  The IC phase viewed from the II limit, in which $t_\rho=0$ and $t_\sigma$ is large.
 The valleys have periodic potential
 $\tilde t_\rho \tau^x \cos\tilde\theta_\rho$ with $\tau^x=\pm 1$, whose
 sign is switched by instanton processes $\tilde v_\sigma$.  }
 \label{vminima}
\end{figure}

Another way to understand this duality, which will prove useful below, is to
consider an instanton expansion for strong coupling.  For large $v_e$
$(\theta_\rho,\theta_\sigma)$ will be tightly bound at the minima of
$V(\theta_\rho,\theta_\sigma)$, shown in Fig. \ref{vminima}(a).  (Here we assume for
simplicity $\eta_\rho=0$.)  The partition function describing the path integral
of (\ref{scc}) can then be expanded in instanton processes, in which $(\theta_\rho,\theta_\sigma)$
switches between nearby minima at discrete times.  Evaluating the first term in (\ref{scc}) for a configuration of
instantons leads to an interaction between the instantons which depends logarithmically
on time.  The expansion describes the partition
function for a one dimensional ``Coulomb gas", where the ``charges" correspond
to the tunneling events.  This Coulomb gas has exactly the same form as the
expansion of (\ref{sii}) in powers of $t_e$, $t_\rho$ and $t_\sigma$.  Thus, we can
identify $t_e$, $t_\rho$ and $t_\sigma$ as the fugacity of the instantons.

This duality argument also works in reverse.  Starting from (\ref{sii}) we can derive (\ref{scc})
by considering large $t_e$ and expanding in instantons in $\tilde\theta_\rho$
and $\tilde\theta_\sigma$ connecting minima in Fig. \ref{vminima}(b), which
have fugacities $v_e$, $v_\rho$ and
$v_\sigma$.

\subsubsection{Charge Insulator/Spin Conductor (IC)}

We next study the mixed charge insulator spin conductor phase.  To generate the
effective action for this phase, including the leading relevant operators
 it is useful to use the instanton analysis discussed at the end of the previous section.
Consider (\ref{sii},\ref{vii}) for large $v_\rho$, keeping $v_e$ and $v_\sigma$ small.
$\theta_\rho$ will be pinned in the minima of $-\cos 2\theta_\rho$,
$\theta_\rho = n \pi$,
while $\theta_\sigma$ remains free to fluctuate.
$(\theta_\rho,\theta_\sigma)$ are thus confined to ``valleys" along the vertical
lines in Fig. \ref{vminima}(c).

There are two types of perturbations to
be considered.  First, $v_e$ will lead to a periodic
potential along the vertical lines, with minima at the dots.  Note,
however, that on alternate lines the sign of the periodic potential
changes, since $\cos \theta_\rho \cos\theta_\sigma  \sim (-1)^n
\cos\theta_\sigma$ for $\theta_\rho = n \pi$.

Next consider an instanton process where
$\theta_\rho$ tunnels between neighboring valleys.  In this process,
$\theta_\rho \rightarrow \theta_\rho \pm \pi$, but $\theta_\sigma$ is
unchanged.  It follows that the $v_e$ perturbation discussed above
changes sign.  Thus, the instanton process does not commute with the $v_e$ term.

The expansion of the partition function in {\it both} instantons and
$v_e$ can be generated by the action for the IC phase give by
$S_{IC}=S_{IC}^0 + S_{IC}^1$
with
\begin{equation}
\label{s0ic}
S^0_{IC} = {1\over\beta}\sum_{\omega_n}{g_\rho \over {2\pi }} |\omega_n|
|\tilde\theta_\rho(\omega_n)|^2 + {1\over {2\pi g_\sigma}} |\omega_n|
|\theta_\sigma(\omega_n)|^2,
\end{equation}
and
\begin{equation}
\label{s1ic}
S^1_{IC} = \int_0^\beta {d\tau\over\tau_c} \left[\tilde t_\rho \tau^x \cos\tilde\theta_\rho
+  \tilde v_\sigma \tau^z \cos \theta_\sigma\right] .
\end{equation}
Here $\tilde t_\rho$ describes the instanton tunneling process.  The
tilde distinguishes it from the ordinary charge tunneling process,
which involves charge $2e$.   $\tilde t_\rho$ describes a tunneling
of charge $e$ {\it without} spin.  $\tilde v_\sigma$ describes the
periodic potential as a function of $\theta_\sigma$ generated by
$v_e$.  We have introduced a pseudo spin degree of freedom $\tau^z =
\pm 1$ to account for the sign of $\cos\theta_\rho$ in the
different valleys.  Since the instanton process switches the sign, it
is associated with $\tau^x$.  Expanding the partition function
defined by (\ref{s0ic},\ref{s1ic}) in powers of $\tilde t_\rho$ and $\tilde v_\sigma$
precisely generates the expansion of (\ref{scc},\ref{vcc}) in instantons.

It is also instructive to derive (\ref{s0ic},\ref{s1ic}) starting from the opposite limit
of the II phase described by (\ref{sii},\ref{vii}).  In this case, consider large
$t_\sigma$, which leads to the horizontal valleys as a function of
$\tilde \theta_\rho$ and $\tilde \theta_\sigma$ in Fig. \ref{vminima}(d).  The
roles of the two terms in (\ref{s1ic}) are thus reversed.  $\tilde t_\rho$
describes the periodic potential along the valleys, which has a sign
specified by $\tau^x = \pm 1$.  $\tilde v_\sigma$ describes the
instanton processes which switch the sign of $\tau^x$.

The lowest order renormalization group flows depend only on the
scaling dimensions of $\tilde t_\rho$ and $\tilde v_\sigma$, and are
unaffected by the pseudospin $\tau^{x,z}$.  We find
\begin{eqnarray}
\label{icdelta}
\Delta(\tilde t_\rho) &=& {1 \over 2 g_\rho}  = {1\over{2g}}  \nonumber\\
\Delta(\tilde v_\sigma) &=& {g_\sigma\over 2} = {1\over{2g}}.
\end{eqnarray}
Thus, the IC phase is stable when $g< 1/2$.

In section III.D we will
require the renormalization group flow to third order in $\tilde
t_\rho$ and $\tilde v_\sigma$.  There, the non trivial interaction
between them introduced by the pseudospin will play a crucial role.

The conductivity at the IC fixed point is given by
\begin{equation}
\label{gabic}
G_{XX} = G_{YY} = G_{XY} = 0.
\end{equation}
This, however, does {\it not} mean that the full four terminal
conductance is zero because spin conservation still requires
$G_{ZZ} = 2 e^2/h$.  This leads to the non trivial structure in the
four terminal conductance predicted in Ref. \onlinecite{chamon1}.

At finite temperature, there will be corrections to the conductance.
We find
\begin{eqnarray}
\label{icdeltagxx}
\delta G_{XX} &=& d_1 \tilde t_\rho^2 T^{g^{-1}-2}  \nonumber \\
\delta G_{YY} &=& d_2 \tilde v_\sigma^2 T^{g^{-1}-2}.
\end{eqnarray}
As in section the corrections to $G_{XY}$ will depend on a higher
order irrelevant operator.  For instance, $\lambda_1  \tau^y \sin
\tilde\theta_\rho \sin\theta_\sigma$ and
$\lambda_2  \tau^y \cos
\tilde\theta_\rho \cos\theta_\sigma$ lead  to
\begin{equation}
\label{icdeltagxy}
\delta G_{XY} = d_3 \lambda_1\lambda_2 T^{2g^{-1} -2}.
\end{equation}
As in (\ref{ccdeltagxy}),
$G_{XY}$ is suppressed more strongly at low temperature than $G_{XX}$
and $G_{YY}$, and the exponent is larger than 2 for $g<1/2$.

\subsubsection{Charge conductor/Spin insulator (CI)}

For $g>2$ the perturbation $v_\sigma \cos 2\theta_\sigma$ in (\ref{vcc})
becomes relevant and drives the system to the CI phase.  This may be
described in a manner similar to the IC phase.  It is described by
the action $S_{CI} = S^0_{CI} + S^1_{CI}$ with
\begin{equation}
\label{s0ci}
S^0_{CI} = {1\over\beta}\sum_{\omega_n}{g_\sigma \over {2\pi }} |\omega_n|
|\tilde\theta_\sigma(\omega_n)|^2 + {1\over {2\pi g_\rho}} |\omega_n|
|\theta_\rho(\omega_n)|^2
\end{equation}
and
\begin{equation}
\label{s1ci}
S^1_{CI} = \int_0^\beta {d\tau\over\tau_c}\left[ \tilde t_\sigma \tau^z \cos(\tilde\theta_\sigma+\eta_\sigma)
+  \tilde v_\rho \tau^x \cos(\theta_\rho + \eta_\rho)\right].
\end{equation}
The leading relevant operators have dimensions
\begin{eqnarray}
\label{deltaci}
\Delta(\tilde t_\sigma) &=& {1\over {2g_\sigma}} = {g\over 2} \nonumber\\
\Delta(\tilde v_\rho) &=& {g_\rho\over 2} = {g\over 2}.
\end{eqnarray}
This phase is thus stable when $g>2$ and has conductance
\begin{eqnarray}
\label{gabci}
G_{XX} &=& G_{YY} = 2 e^2/h \nonumber\\
G_{XY} &=& 0.
\end{eqnarray}

\subsubsection{Spin orbit interactions, and the T-Breaking Insulator}

In this section we consider the role of spin orbit interaction terms
which violate the conservation of spin $S_z$, but respect time
reversal symmetry.  We will argue that such terms are irrelevant for
the critical behavior of the point contact when $g>1/2$, but they are
relevant for $g<1/2$ and drive the system at low energy to a time
reversal symmetry breaking insulator (TBI).

\begin{figure}
\centerline{ \epsfig{figure=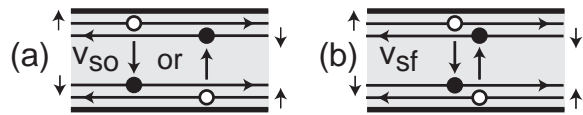,width=3in} }
 \caption{Tunneling processes in the CC limit allowed by spin nonconserving
 spin orbit interactions.  $v_{so}$ is a single particle process where
 a single spin is flipped, while
 $v_{sf}$ is a two particle process, flipping two spins.}
 \label{sotunneling}
\end{figure}

Time reversal symmetry allows the following terms in the expansion
about the CC fixed point (\ref{scc}).
\begin{equation}
\label{sso}
S_{CC}^{SO} = \int_0^\beta {d\tau\over\tau_c} \left[v_{so} \cos\varphi_\sigma \sin\theta_\sigma
+ v_{sf} \cos(2 \varphi_\sigma+\eta_{sf})\right].
\end{equation}
The first term is a single electron process $\psi_{R\uparrow}^\dagger\psi_{R\downarrow}$
(Fig. \ref{sotunneling}(a)) in which an electron
flips its spin and crosses the junction.  The second term is a
correlated tunneling process $\psi_{R\uparrow}^\dagger \psi_{L\uparrow}^\dagger
\psi_{R\downarrow}\psi_{L\downarrow}$(Fig. \ref{sotunneling}(b)),
where a left and right moving pair of up spins
flip into a left and right moving pair of down spins.  Referring to Table \ref{symtab}, it
is clear that both terms respect time reversal symmetry.  $\eta_{sf}$ is allowed by
time reversal symmetry, but violates both mirrors ${\cal M}_x$ and ${\cal M}_y$.  Higher
order processes are also possible, though they will be less relevant
perturbatively.

It is straightforward to determine the scaling dimensions of these
perturbations.  We find,
\begin{eqnarray}
\label{vso}
\Delta(v_{so}) &=& {1\over 2}(g_\sigma + g_\sigma^{-1}) = {1\over 2}(g+g^{-1})\nonumber\\
\Delta(v_{sf}) &=& {2\over g_\sigma} = 2 g.
\end{eqnarray}
For $g\ne 1$ the single particle spin orbit term, $v_{so}$ is
always irrelevant.  However, $v_{sf}$ becomes relevant when $g<1/2$.

At finite temperature these lead to corrections to the conductance of the CC
phase.  To lowest order they do not affect $G_{XX}$, $G_{XY}$
and $G_{YY}$.  However we find
\begin{equation}
\label{gso}
\delta G_{ZZ} \propto   \left\{\begin{array}{ll}
 T^{g + g^{-1}-2} & g>1/\sqrt{3} \\
 T^{4g-2} & g < 1/\sqrt{3}
 \end{array}\right.
\end{equation}
Like $G_{XY}$, $G_{ZX}$ and $G_{ZY}$ are zero unless higher order
irrelevant operators, which involve extra powers of
$\partial_x\varphi_\alpha$ or $\partial_x\theta_\alpha$, are
included.  We find
\begin{eqnarray}
\label{gxzgyz}
\delta G_{ZX} &\propto & T^{2g} \\
\delta G_{ZY} &\propto& T^{g + g^{-1}}.
 \nonumber
\end{eqnarray}
For weak interactions, $g\sim 1$ these conductances vanish for
$T\rightarrow 0$ as $T^2$.

\begin{figure}
\centerline{ \epsfig{figure=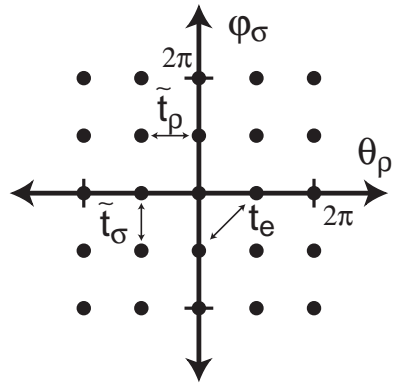,width=2.in} }
 \caption{Minima of the potential $V(\theta_\rho,\varphi_\sigma)$ in
 \ref{s1ccso}.  Large  $v_\rho$ and $v_{so}$ define the time reversal breaking
 insulating phase.  Instanton processes $\tilde t_\rho$ and $\tilde t_\sigma$ restore
 time reversal invariance.  They correspond to tunneling of spinless
 chargons or chargeless spinons.    }
 \label{sominima}
\end{figure}

For $g<1/2$ there are two relevant perturbations about the CC limit.
To study their effects we consider a model in which only the relevant
perturbations appear.  Since these perturbations involve the
commuting operators
$\varphi_\sigma$ and $\theta_\rho$, it is useful to study the $0+1$
dimensional field theory of those variables
\begin{equation}
\label{s0ccso}
S^0_{CC} = \sum_{\omega_n}{1 \over {2\pi g_\rho}} |\omega_n|
|\theta_\rho(\omega_n)|^2 + {g_\sigma\over {2\pi}} |\omega_n|
|\varphi_\sigma(\omega_n)|^2,
\end{equation}
with
\begin{equation}
\label{s1ccso}
S^1_{CC} =  \int_0^\beta {d\tau\over\tau_c}\left[ v_\rho \cos(2\theta_\rho+\eta_\rho)
+  v_{sf} \cos(2\varphi_\sigma+\eta_{sf})\right].
\end{equation}
The low temperature behavior of this theory can be studied by the
duality arguments of section II.B.2.   When $v_{\rho}$ and $v_{sf}$ are
both large, $(\theta_\rho,\varphi_\sigma)$ will be stuck in the deep
minima of $V_{CC}(\theta_\rho,\varphi_\sigma)$ shown in Fig. \ref{sominima}.  In
this phase, the {\it four terminal} conductance is zero,
\begin{equation}
\label{gtbi}
G_{AB} = 0.
\end{equation}
This can be seen most simply by renaming the variables
\begin{eqnarray}
\label{theta12tbi}
\theta_\rho &\rightarrow &\theta_1 + \theta_2 \nonumber\\
\varphi_\sigma &\rightarrow &\theta_1 - \theta_2 \nonumber\\
\varphi_\rho &\rightarrow &\varphi_1 + \varphi_2 \\
\theta_\sigma &\rightarrow &\varphi_1 - \varphi_2. \nonumber
\end{eqnarray}
The interpretation of $\theta_{1(2)}$ and $\varphi_{1(2)}$ is
simple.  They are the usual Luttinger liquid charge and phase
variables for the top (bottom)
edges in Fig. \ref{tunnelingop}(a,b,c).  In the strong coupling phase $\theta_1$ and
$\theta_2$ are both pinned, so that any current flowing in from any
lead is perfectly reflected back into that lead.  The four leads are
completely decoupled.

This is the same perfectly reflecting phase that would arise if we had a single
particle backscattering term on each edge
$v_{\rm back}(\cos 2\theta_1 + \cos 2\theta_2)= 2v_{\rm back}\cos\theta_\rho \cos \varphi_\sigma$,
which would be relevant for $g<1$.
However in our problem that term is forbidden by time reversal
symmetry. It is thus clear that time reversal symmetry is violated by
the strong coupling fixed point.  It is useful to see this from Fig.
\ref{sominima}.  Note that since under time reversal $\varphi_\sigma \rightarrow
\varphi_\sigma + \pi$.  Thus pinning $\varphi_\sigma$ violates time reversal.
There are two sets of minima of
$V(\theta_\rho,\varphi_\sigma)$ which are interchanged by the time reversal
operation.

At finite temperature tunneling processes between the two sets of
minima of
$V(\theta_\rho,\varphi_\sigma)$ will restore time reversal symmetry.
These instanton processes correspond to tunneling of charge from one
lead to another.  Interestingly, the lowest order instanton
processes, denoted $\tilde t_\rho$ and $\tilde t_\sigma$ do {\it not}
correspond to tunneling of electrons, but rather spinless charge $e$
``chargons", or charge neutral ``spinons".

The scaling dimensions of these instanton processes can be deduced
from (\ref{s0ccso},\ref{s1ccso}).  We find
\begin{eqnarray}
\label{sodelta}
\Delta(\tilde t_\rho) &=& {1\over{2 g_\rho}} = {1\over {2g}} \nonumber\\
\Delta(\tilde t_\sigma) &=& {g_\sigma\over 2} = {1\over {2g}}.
\end{eqnarray}
Thus, both processes are irrelevant for $g<1/2$, and the TBI phase is
stable.  These processes lead to power law temperature behavior,
\begin{eqnarray}
\label{deltagxxtbi}
\delta G_{XX} &=& c_1 \tilde t_{\rho}^2 T^{1/g - 2} \nonumber\\
\delta G_{YY} &=& c_1 \tilde t_{\sigma}^2 T^{1/g - 2}.
\end{eqnarray}

When the $\tilde t_{\rho,\sigma}$ processes dominate, there will be non trivial
noise correlations in the current.  The $\tilde t_\rho$ process involves
transferring charge $e/2$ from lead 1 to lead 2 and another $e/2$ from lead 4
to lead 3.  This leads to correlations in the low frequency noise defined by
\begin{equation}
\label{noisetbi}
{\cal S}_{ij}(\omega) = \int dt e^{i\omega t} \langle I_i(t) I_j(0) + I_j(0) I_i(t)
\rangle.
\end{equation}
Consider
the two terminal geometry $I_X = G_{XX} V_X$.  The current $I_X$ will be
carried by the $\tilde t_\rho$ processes, so that $I_1 = I_4 = I_X/2$.  The
shot noise correlations in the limit $\omega\rightarrow 0$ will be
\begin{equation}
S_{11} = S_{44} = S_{14} = S_{41} = 2 e^* I_1
\end{equation}
with $e^* = e/2$.  Thus, the currents are all perfectly correlated, and
the current in each lead is carried by {\it
fractional} charges, $e/2$.


\section{Critical behavior of conductance}

In this section we describe the critical behavior of the conductance at the
pinch-off transition of the point contact.  We will compute the critical
conductance $G_g^*$, the critical exponent $\alpha_g$ and the scaling function
${\cal G}_g(X)$ in certain solvable limits.  We will begin in section IIIA with
a discussion of the general properties of the scaling function and a summary of
our calculated results.  Then in the following sections we will describe in
detail our calculations for $g=1-\epsilon$, $g=1/\sqrt{3}$ and $g=1/2 +
\epsilon$.

\subsection{Scaling behavior and summary of results}

\begin{figure}
\centerline{ \epsfig{figure=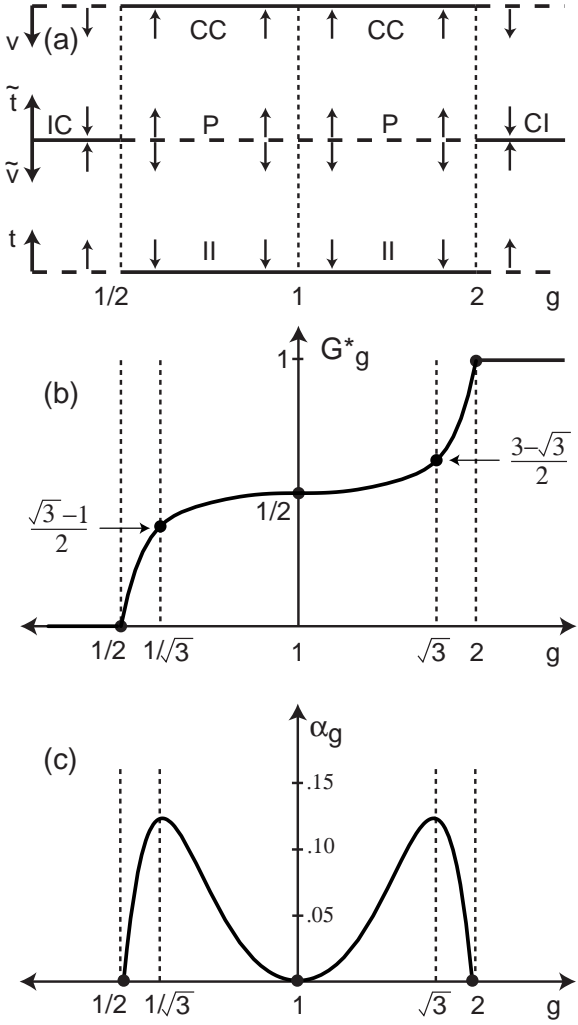,width=3.in} }
 \caption{(a) Phase diagram for a point contact in a QSHI
 as a function of the Luttinger liquid parameter $g$.  The arrows indicate the
 stability of the CC, II, CI and IC phases, as well as the critical fixed point P.
 This figure assumes spin conservation.
 In the presence of spin orbit interactions, the IC phase is unstable for $g<1/2$.
 This leads to the TBI phase discussed in section II.B.5.  (b) Conductance $G^*$
 of the critical fixed point as a function of $g$.  The curve is a
 fit, which incorporates the data in
 (\ref{gstar}).  (c) Critical exponent $\alpha_g$ as a function of $g$.
 The curve is a fit incorporating the data in (\ref{alphagg}).  $g$ is plotted
 on a log scale in all three panels to emphasize the $g \leftrightarrow 1/g$
 symmetry.
  }
 \label{phasediagram}
\end{figure}

The stability analysis of the previous sections leads to the phase
diagram as a function of $g$ depicted in Fig. \ref{phasediagram}(a).  The top line
depicts the CC phase and the bottom line depicts the II phase, and
the arrows denote the stability associated with the leading relevant
operators.  Since the II and CC phases are both stable for $1/2<g<2$
they are separated by an intermediate unstable fixed point P,
denoted by the dashed central line.  For $g<1/2$ the II and CC phases
become unstable, and when spin is conserved the flow is toward the IC
phase.  We will see in section III.D that the unstable critical fixed point
matches smoothly onto the IC fixed point at $g=1/2$.  Similarly, the
CI fixed point is stable for $g>2$, and connects to the critical
fixed point at $g=2$.

For $1/2<g<2$ the unstable intermediate fixed point P describes
the critical behavior of the pinch-off transition of the
point contact.  We will argue that this fixed point is
characterized by a {\it single} relevant operator, which allows us to
formulate a single parameter scaling theory for the pinch-off
transition.  If we denote $u$ as the relevant operator, then the
leading order renormalization group flow near the fixed point has the
form,
\begin{equation}
\label{dudl}
du/d\ell = \alpha_g u,
\end{equation}
where $\alpha_g$ is a critical exponent to be determined.
By varying a gate voltage $V_G$ it is possible to cross
from the region of stability of the II phase to the region of
stability of the CC phase.  In the process one must pass through the
fixed point $u=0$ at $V_G=V_G^*$.  Near the transition, we thus have
$u \propto \Delta V_G = V_G - V_G^*$.  Under a renormalization group
transformation in which energies length and time are rescaled by $b$,
we have $u \rightarrow u b^{\alpha_g}$ and $T \rightarrow T b$.  Invariance under this
transformations requires that physical quantities can only depend on
$u$ and $T$ in the combination $u/T^{\alpha_g}$.
Close to the transition we thus have
\begin{equation}
\label{gabscaling}
\lim_{T,\Delta V_G \rightarrow 0}  G_{AB}(T,\Delta V_G) =
2 {e^2\over h} {\cal G}_{g,AB}(c {\Delta V_G \over T^{\alpha_g}}),
\end{equation}
where $c$ is a nonuniversal constant
and ${\cal G}_{g,AB}$ is a universal crossover scaling function which varies
between $0$ and $1$.

We will argue that the critical point characterizing the pinch-off transition
has emergent spin conservation as well as mirror symmetry, so that
the only nonzero elements of the conductance matrix are $G_{XX}$ and
$G_{YY}$.  Moreover, the duality considerations discussed in section
III.C require that ${\cal G}_{g,YY}(X)$ and ${\cal G}_{g,XX}(X)$ are related, so that
they are both determined by the same universal scaling function,
\begin{eqnarray}
\label{gsymmetry-x}
{\cal G}_{g,XX}(X) &=& {\cal G}_g(X), \nonumber\\
{\cal G}_{g,YY}(X) &=& {\cal G}_g(-X).
\end{eqnarray}

The scaling function ${\cal G}_g(X)$ has some general properties which are easy
do deduce.  First, the equivalence between the CC theory
at $g$ with the II theory at $1/g$ leads to the relation
\begin{equation}
\label{gsymmetry1/g}
{\cal G}_{1/g}(X) = 1 - {\cal G}_g(-X).
\end{equation}
Second, when $T \rightarrow 0$ for fixed $\Delta V_G$ the system
flows to either the CC or the II phase, where the temperature dependence of the
conductance is given by (\ref{ccdeltag}).  The behavior of the scaling function for large $X$
then follows,
\begin{eqnarray}
\label{gasymptote}
{\cal G}_g(X \rightarrow +\infty)  &=& 1 - a^+_g X^{-\beta^+_g} \nonumber \\
{\cal G}_g(X \rightarrow -\infty)  &=& a^-_g X^{-\beta^-_g}.
\end{eqnarray}
The coefficients
$a_g^\pm$ depend on the normalization of $X$, but can be fixed if we
specify ${\cal G}_g'(X=0) = 1/2$.  The exponents obey the relations
\begin{eqnarray}
\beta^+_g &=& \left\{\begin{array}{ll}
(4g-2)/\alpha_g &  1/2 < g < 1/\sqrt{3}\nonumber \\
(g+g^{-1}-2)/\alpha_g  &  1/\sqrt{3}<g < 1
\end{array}\right. \\
\label{beta}
\beta^-_g &=& (g+g^{-1}-2)/\alpha_g \quad\quad 1/2<g<1.
\end{eqnarray}
The behavior of $\beta^\pm_g$ for $1<g<2$ can be deduced using (\ref{gsymmetry1/g}).

In the following sections we compute properties of the scaling
function at $g=1-\epsilon$, $g=1/\sqrt{3}$ and $g = 1/2+\epsilon$.
From (\ref{gsymmetry1/g}) we can deduce corresponding results at $g=1+\epsilon$,
$g=\sqrt{3}$ and $g=2-\epsilon$.  First consider the critical
conductance $G_g^* = {\cal G}_g(X=0)$.  We find,
\begin{equation}
\label{gstar}
G_g^* = \left\{\begin{array}{ll}
1/2 + O(\epsilon^3) &  g = 1 - \epsilon \\
(\sqrt{3}-1)/2  &   g = 1/\sqrt{3} \\
\pi^2 \epsilon &   g = 1/2 + \epsilon . \\
\end{array}\right.
\end{equation}
These results are summarized in Fig. \ref{phasediagram}(b).  The curve is a
polynomial fit of $G^*(\log g)$ which incorporates the data in Eq.
(\ref{gstar}) and the $g\leftrightarrow 1/g$ symmetry.  It is satisfying that
the curve is smooth and monotonic, which indicates a
consistency between the slopes at $g=1/2,1$ and the value at
$g=1/\sqrt{3}$.

We are able to deduce the critical exponent $\alpha_g$ for $g=1 -
\epsilon$ and $g= 1/2+\epsilon$.  We find
\begin{equation}
\label{alphagg}
\alpha_g = \left\{\begin{array}{ll}
\epsilon^2/2 & g = 1 - \epsilon \\
4 \epsilon & g = 1/2 + \epsilon. \\
\end{array}\right.
\end{equation}
These results are summarized in Fig \ref{phasediagram}(c).  The curve is a polynomial
fit of $\alpha(\log g)$.  It is suggestive that in this fit
$\alpha_g$ exhibits a maximum near $g=1/\sqrt{3}$ with a value
$\alpha_{1/\sqrt{3}}=.123 \sim 1/8$.  It is possible, however, that
$\alpha_g$ exhibits a cusp at $g=1/\sqrt{3}$ analogous to the
behavior of $\beta_g$ in (\ref{beta}).

\begin{figure}
\centerline{ \epsfig{figure=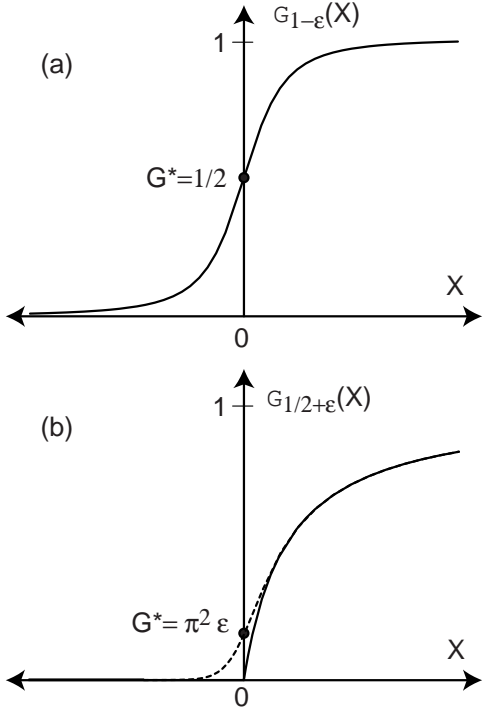,width=2.5in} }
 \caption{The universal scaling function ${\cal G}_g(X)$ for (a) $g=1-\epsilon$
 (Eq. \ref{g1-epsilon}) and (b) $g=1/2+\epsilon$ (Eq. \ref{g1/2+epsilon}).
 In (b) the solid line is $\epsilon\rightarrow 0$, and the dashed line shows
 the approximate behavior for $\epsilon\sim .02$.  }
 \label{scalingfig}
\end{figure}

In sections III.C and III.D we compute the full scaling function ${\cal G}_g(X)$ in the
limits $g= 1-\epsilon$ and $g = 1/2 + \epsilon$ to lowest order in
$\epsilon$.  For $g=1+\epsilon$, $\epsilon\rightarrow 0$ we find
\begin{equation}
\label{g1-epsilon}
{\cal G}_{1}(X) = {1\over 2}\left(1 + {X \over
\sqrt{1+X^2}}\right).
\end{equation}
For $g=1/2+\epsilon$, $\epsilon\rightarrow 0 $
\begin{equation}
 \label{g1/2+epsilon}
{\cal G}_{1/2}(X) = \theta(X) {X\over{1+X}}.
\end{equation}
The singular behavior near $X=0$ in (\ref{g1/2+epsilon}) is rounded
for finite $\epsilon$.  The perturbative analysis in Section III.D.1 shows
that for $|X|\ll 1$
\begin{equation}
\label{g1/2+epsilon2}
{\cal G}_{1/2+\epsilon}(X) = {X\over{1-e^{-X/(\pi^2\epsilon)}}}.
\end{equation}
${\cal G}_{1-\epsilon}(X)$ and ${\cal G}_{1/2+\epsilon}(X)$
are plotted in Figs. \ref{scalingfig}(a) and \ref{scalingfig}(b).  For $g$ close to
$1$ the pinch-off curve is symmetrical about $G^* = e^2/h$.  However, for
stronger repulsive interactions it becomes asymmetrical, as $G^*$ is
reduced, approaching $0$ at $g=1/2$.

The asymptotic $|X|\rightarrow \infty$ behavior (\ref{gasymptote}) of
${\cal G}_1(X)$ and ${\cal G}_{1/2+\epsilon}$ can also be
determined from (\ref{g1-epsilon},\ref{g1/2+epsilon}), though a
separate calculation (see III.D.3) is required for
${\cal G}_{1/2+\epsilon}(X\rightarrow
-\infty)$.  The results, which are consistent with (\ref{beta})
are shown in Table \ref{asymptable}.

\begin{table}
\centering
\begin{tabular}{|c|c|c|c|c|}\hline
$g$ & $\beta_g^+$ & $a_g^+$ & $\beta_g^-$ & $a_g^-$\\
\hline
$1-\epsilon$ & $2$ & $1/4$ & $2$ & $1/4$ \\
$1/2+\epsilon$ & $1$& $1$ & $1/(8\epsilon)$ &
$(2.75)^{1/(8\epsilon)}$
\\
  \hline
\end{tabular}
  \caption{Parameters in Eq. \ref{gasymptote} for the
  asymptotic behavior of the scaling function ${\cal G}_g(X)$ in the
  solvable limits $g\rightarrow 1$, $g\rightarrow 1/2$.
}
  \label{asymptable}
\end{table}

\subsection{Quantum Brownian Motion Model, Duality and $g = 1/\sqrt{3}$}

In this section we recast the Luttinger liquid model as a model of QBM
in a periodic potential.
This mapping elucidates the duality between the CC and II limits and
exposes an extra
symmetry the problem at $g=1/\sqrt{3}$ which allows us to deduce the
critical conductance at that point.  We begin with a brief review of the QBM
model and then derive its consequences for the scaling function ${\cal G}_g(X)$
and $G^*_{1/\sqrt{3}}$.

\subsubsection{Quantum Brownian Motion Model}

The QBM model\cite{schmid,fz,guinea} was
originally formulated as a theory of the motion of a heavy particle coupled to
an Ohmic dissipative environment modeled as a set of Caldeira Leggett
oscillators\cite{caldeira}.  Though the applicability of this model to the motion of a real
particle coupled to phonons or electron-hole pairs has been questioned\cite{itai,zimanyi}, it was
later shown that this model is directly relevant to quantum impurity problems.
Specifically, the QBM model in a
one dimensional periodic potential is equivalent to the theory of a weak link
in a single channel Luttinger liquid\cite{kf1,kf2}.  In this mapping the QBM
 takes place in an abstract space where the ``coordinate" of the
``particle" is the number of electrons that have tunneled past the weak link.
The periodic potential is due to the discreteness of the electron's charge.
The low energy excitations of the Luttinger liquid play the role of the
dissipative bath, and the strength of the dissipation is related to the
Luttinger liquid parameter $g$.
The one dimensional QBM model has two phases: a localized phase with
conductance $G=0$ stable for $g<1$ and a fully coherent phase with perfect conductance
stable for $g>1$.

The SLL model corresponds to a QBM
model in a two dimensional periodic potential, where the ``coordinates" are the
spin and charge variables $\theta_{\rho,\sigma}$.  This model is richer than its
one dimensional counterpart because it admits additional fixed
points which are intermediate between localized and perfect.  These fixed
points were first found in the Luttinger liquid model\cite{furusaki,kf1}, and later formulated in
terms of the QBM\cite{yikane}.  For certain values of $g_\rho$ and
$g_\sigma$ these intermediate fixed points are related to the 3 channel Kondo
problem\cite{yikane} and the 3 state Potts models\cite{affleck1}.  However, those limits are not directly
applicable to the QSHI model, where $g_\rho = 1/g_\sigma = g$.  We
will show that when $g=1/\sqrt{3}$ the critical fixed point of the QSHI point
contact corresponds to the
intermediate point discussed in Ref. \onlinecite{yikane} for a QBM model on a triangular
lattice.

To formulate the QBM model we begin with the
action (\ref{scc}) and define new rescaled variables,
\begin{equation}
\theta_\alpha = \pi \sqrt{2g_\alpha}r_\alpha.
\end{equation}
Then (\ref{scc}) takes the form,
\begin{equation}
\label{sqbmr}
S ={1\over {4\pi\beta}}\sum_n |\omega_n| |{\bf r}(\omega_n)|^2 - \int {d\tau\over{\tau_c}} \sum_{\bf G}
v_{\bf G} e^{2\pi i {\bf G}\cdot {\bf r}(\tau)}.
\end{equation}
The periodic potential is
characterized by reciprocal lattice vectors ${\bf G}=m_1 {\bf b}_1 + m_2 {\bf b}_2$.
The primitive reciprocal lattice vectors ${\bf b}_{1,2}$ correspond to the single electron back
scattering processes, and are given by
\begin{equation}
\label{b1b2}
{\bf b}_1 = {1\over\sqrt{2}}( \sqrt{g_\rho}, \sqrt{g_\sigma});\quad
{\bf b}_2 = {1\over\sqrt{2}}(\sqrt{g_\rho},-\sqrt{g_\sigma}).
\end{equation}
The Fourier components of the periodic potential are
$v_{{\bf b}_1} = v_{{\bf b}_2} = v_e e^{i\eta_\rho}/4$,
$v_{{\bf b}_1 + {\bf b}_2} = v_\rho/2$ and
$v_{{\bf b}_1 - {\bf b}_2} = v_{\sigma}/2$.

The dual theory is obtained by expanding the partition function for large $v_{\bf G}$ in powers
of instantons.  When $v_{\bf G}$ is large, the potential has minima on a real space lattice
${\bf R} = n_1 {\bf a}_1 + n_2{\bf a}_2$.  The primitive lattice vectors
satisfy ${\bf a}_i \cdot {\bf b}_j = \delta_{ij}$ and are given by
\begin{equation}
\label{a1a2}
{\bf a}_1 = {1\over\sqrt{2}}( {1\over\sqrt{g_\rho}},{1\over\sqrt{g_\sigma}}); \quad
{\bf a}_2 = {1\over\sqrt{2}}( {1\over\sqrt{g_\rho}},-{1\over\sqrt{g_\sigma}}).
\end{equation}
The expansion in instantons connecting these minima is generated by the action
\begin{equation}
\label{sqbmk}
S ={1\over{4\pi\beta}}\sum_n |\omega_n| |{\bf k}(\omega_n)|^2 - \sum_{\bf R}
\int {d\tau\over{\tau_c}} t_{\bf R} e^{2\pi i {\bf R}\cdot {\bf k}(\tau)}.
\end{equation}
This is equivalent to (\ref{sii},\ref{vii}) with $k_\alpha = \pi
\sqrt{g_\alpha/2}\tilde\theta_\alpha$ and
$t_{{\bf a}_1} = t_{{\bf a}_2} = t_e e^{i\eta_\rho}/4$,
$t_{{\bf a}_1 + {\bf a}_2} = t_\rho/2$,
$t_{{\bf a}_1 - {\bf a}_2} = t_{\sigma}/2$.

With the above normalizations for ${\bf r}$ and ${\bf k}$ the scaling
dimensions of the potential perturbations are
\begin{equation}
\label{deltaqbm}
\Delta(v_{\bf G}) = |{\bf G}|^2; \quad
\Delta(t_{\bf R}) = |{\bf R}|^2.
\end{equation}
Since operators are relevant when $\Delta<1$,
the most relevant potentials are those with the smallest lattice (reciprocal
lattice) vectors $|{\bf R}_{\rm min}|$ ($|{\bf G}_{\rm min}|$).  As shown in Refs.
\onlinecite{kf1} and \onlinecite{yikane}
there are ranges of $g_\rho$ and $g_\sigma$ where both $|{\bf R}_{\rm min}|$ and
$|{\bf G}_{\rm min}| >1$, so that both phases are perturbatively stable.  An
unstable intermediate fixed point must therefore be present between them.

\begin{figure}
\centerline{ \epsfig{figure=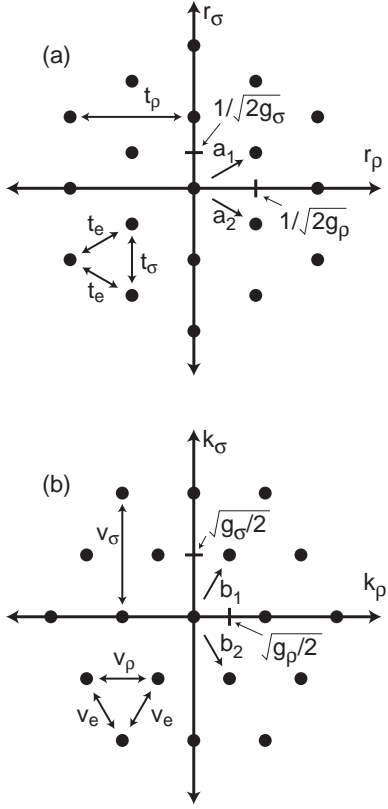,width=2in} }
 \caption{(a) Minimia of the periodic potential $V({\bf r})$ in (\ref{sqbmr}).
 (b) Minima of $V({\bf k})$ in the dual theory (\ref{sqbmk}).  When
 $g_\sigma = 3 g_\rho$ both periodic
 potentials have triangular symmetry at the critical point, which
 implies the mobility $\mu^*_{\alpha\beta}$
 is isotropic.  This occurs at $g=1/\sqrt{3}$. }
 \label{hexagons}
\end{figure}

This
fixed point can be accessed perturbatively when $|{\bf R}_{\rm min}|$ and $|{\bf
G}_{\rm min}|$ are close to 1.  While this does {\it not} occur in the regime
$g_\rho = 1/g_\sigma$ relevant to the QSHI problem, it is instructive to study
this perturbation theory because it provides evidence that the critical fixed
point has emergent mirror and spin conservation symmetry.

When $g_\rho = 1/2 + \epsilon_\rho$ and $g_\sigma = 3/2 + \epsilon_\sigma$ the
period potential has triangular symmetry, which is slightly distorted if
$\epsilon_\sigma \ne 3 \epsilon_\rho$.  If we denote the relevant variables as
$v_1 = v_{{\bf b}_1} = v_{{\bf b}_2} = v_e e^{i\eta_\rho}/4$ and
$v_2 = v_{{\bf b}_1+{\bf b}_2} =v_\rho/2$, the second order renormalization group
flow equations are\cite{kf1}
\begin{eqnarray}
\label{dv12dl}
dv_1/d\ell &=& {1\over 2}(\epsilon_\rho + \epsilon_\sigma)v_1 - 2 v^*_1 v_2 \nonumber \\
dv_2/d\ell &=& 2 \epsilon_\rho v_2 - 2 v_1^2.
\end{eqnarray}
These equations describe an intermediate fixed point with a single unstable
direction at $v_1 = \sqrt{\epsilon_\rho(\epsilon_\rho+\epsilon_\sigma)}/2$ and
$v_2 = (\epsilon_\rho+\epsilon_\sigma)/4$.  Note that at the critical point
$v_1$ is {\it real}, so that $\eta_\rho = 0$.  Thus the critical
point has an emergent mirror symmetry even if the bare parameters in the model
do not.  Moreover, the flow out of the fixed point along the single unstable
direction is also along a line with $v_1$ real.  Thus the crossover between the
intermediate fixed point and the trivial fixed point, which determines the
crossover scaling function also has emergent mirror symmetry.  Mirror symmetry
breaking is an {\it irrelevant} perturbation at the critical fixed point.

If $\epsilon_\sigma= 3\epsilon_\rho$ then the lattice vectors have a triangular
symmetry.  In this case, the fixed point is at $v_1 = v_2 = \epsilon_\rho$.
This means that the periodic potential at the fixed point has emergent {\it triangular
symmetry}, even when the bare potential does not.  The unstable flow out of the
fixed point is also along the high symmetry line $v_1=v_2$.

It seems quite likely that the critical fixed point and unstable flows connecting it
to the trivial fixed points retain their high symmetry even
outside the perturbative small $\epsilon$ regime.  This suggests that in general
the critical fixed point has mirror symmetry, and that at $g=1/\sqrt{3}$ it has
triangular symmetry.  We will use this fact below to determine the critical
conductance at $g=1/\sqrt{3}$.

\subsubsection{Kubo conductance, mobility and duality relations}

The spin and charge conductances  in the Luttinger liquid model computed by the Kubo
formula are given by a retarded current-current correlation function. For the present discussion
it is useful to write this as an imaginary time correlation function, which can
be analytically continued to real time via $i\omega \rightarrow \omega + i
\eta$ before taking the $\omega\rightarrow 0$ limit.  Then
\begin{equation}
\label{kubo}
G^K_{\alpha\beta}(i\omega_n) = {1\over {\hbar |\omega_n|}} \int d\tau e^{i\omega_n \tau}
\langle J_\alpha(\tau) J_\beta(0)\rangle,
\end{equation}
where the spin and charge currents are
$J_\alpha = e\partial_t \theta_\alpha/\pi =  e [\theta,{\cal H}]/(i\pi\hbar)$.
This may be expressed as
\begin{equation}
\label{gkvsmu}
G^K_{\alpha\beta}(\omega_n) = 2 {e^2\over h} \sqrt{g_\alpha g_\beta} \mu_{\alpha\beta}(\omega_n),
\end{equation}
where the {\it mobility} of the QBM model is
\begin{equation}
\label{mur}
\mu_{\alpha\beta}(\omega_n) = 2\pi |\omega_n| \langle r_\alpha(-\omega_n)
r_\beta(\omega_n)\rangle.
\end{equation}
$\mu_{\alpha\beta}$ is normalized so that when $v_{\bf G}=0$
$\mu_{\alpha\beta} = \delta_{\alpha\beta}$.

The conductance -- or equivalently $\mu_{\alpha\beta}$
 can also be computed from the dual model.  It is given by
\begin{equation}
\label{mudual}
\mu_{\alpha\beta} = \delta_{\alpha\beta} - \tilde\mu_{\alpha\beta},
\end{equation}
where the {\it dual} mobility is
\begin{equation}
\label{muk}
\tilde\mu_{\alpha\beta}(\omega_n) = 2\pi |\omega_n| \langle k_\alpha(-\omega_n)
k_\beta(\omega_n)\rangle.
\end{equation}
(\ref{mudual},\ref{muk}) are obvious in the perfectly transmitting and perfectly reflecting limits.
They can be derived more generally by starting with a Hamiltonian formulation of
the action, analogous to (\ref{h0thetaphi}), which involves both ${\bf r}$ and ${\bf k}$.
$\mu_{\alpha\beta}$ can then be computed either by first integrating out ${\bf
k}$ to obtain (\ref{mur}) or first integrating out ${\bf r}$ to obtain (\ref{muk}).

Since $g_\rho = 1/g_\sigma = g$, the dual theory depicted in Fig.
\ref{hexagons}(b)
is identical to the original
theory shown in Fig. \ref{hexagons}(a) with the identification $r_\rho \leftrightarrow k_\sigma$, $r_\sigma
\leftrightarrow k_\rho$.  It follows that the mobility $\mu^*_{\alpha\beta}$ of the fixed
point satisfies
\begin{equation}
\label{mustardual}
\mu^*_{\alpha\beta} = \left[\sigma^x \tilde \mu^*
\sigma^x\right]_{\alpha\beta}.
\end{equation}
In addition, if $u$ parameterizes the relevant direction at the critical fixed point, then under
the duality $u \rightarrow -u$.   It follows that slightly away from the
critical fixed point we have
\begin{equation}
\label{muudual}
\mu_{\alpha\beta}(u) = \left[\sigma^x \tilde \mu(-u)
\sigma^x\right]_{\alpha\beta}.
\end{equation}
Properties (\ref{mudual}) and (\ref{muudual}) imply that
$\mu_{\rho\rho}(u) = 1 - \mu_{\sigma\sigma}(-u)$.  Using
(\ref{gabscaling},\ref{gkvsmu},\ref{gxgrho}), this
leads directly to the property (\ref{gsymmetry-x}) of the crossover
scaling function.

An additional set of relations follows from the equivalence between the theory
characterized by $g$ and the dual theory characterized by $1/g$.  From this we
conclude that
\begin{equation}
\mu_{g, \alpha\beta}(u) = \tilde \mu_{1/g, \alpha\beta}(u).
\end{equation}
This, combined with (\ref{gabscaling},\ref{gkvsmu},\ref{mudual},\ref{gxgrho}))
leads to (\ref{gsymmetry1/g}).

\subsubsection{Conductance at $g=1/\sqrt{3}$.}

When $g=1/\sqrt{3}$ the lattice generated by ${\bf b}_1$ and ${\bf b}_2$ has
triangular symmetry.  In section III.B.1 we argued that this means that at the
critical fixed point the periodic potential also has triangular symmetry.  The
$C_6$ rotational symmetry of the triangular lattice requires that the mobility
is {\it isotropic}:
\begin{equation}
\label{muisotropic}
\mu_{\alpha\beta} = \mu_0 \delta_{\alpha\beta}.
\end{equation}
Combining (\ref{mudual}), (\ref{mustardual}), and (\ref{muisotropic}) requires that
\begin{equation}
\label{mu1/2}
\mu_0 = {1\over 2}.
\end{equation}
It follows from (\ref{gkvsmu}) that the Kubo formula spin and charge conductances are given by
\begin{equation}
\label{gk1/sqrt3}
G^K_{\rho\rho} = \sqrt{3} {e^2\over h}; \quad G^K_{\sigma\sigma} = {1\over\sqrt{3}} {e^2\over h}.
\end{equation}

It is well known that the physical conductance measured with leads is not given
by the Kubo conductance\cite{maslov,safi,ponomarenko,kawabata,chamon2}.  Rather, the Kubo conductance needs to be modified to
account for the contact resistance between the Luttinger liquid and the leads.
In appendix A we review the relation between the physical four terminal conductance and the
Kubo conductance.  From (\ref{gxxgkubo}) we conclude that
\begin{equation}
\label{gxx1/sqrt3}
G_{XX} = G_{YY}= (\sqrt{3}-1){e^2\over h}.
\end{equation}

\subsection{Weak interactions : $g=1-\epsilon$}

In this section we develop a perturbative expansion for weak interactions to
compute exactly the crossover scaling function ${\cal G}_g(X)$ as well as the
critical exponent $\alpha_g$ for $g=1-\epsilon$.  A similar approach was
employed by Matveev, Yue and Glazman\cite{glazman} to compute the scaling function
for the crossover between the weak barrier and strong barrier limits
in a single channel Luttinger liquid.  In the single channel problem the
transmission for non interacting electrons is characterized by a transmission
probability $\mathcal{T}$.  Weak forward scattering interactions lead to an exchange
correction to $\mathcal{T}$ at first order in the interactions.  This correction diverges
for $E \rightarrow E_F$ as $\log |E-E_F|$.  Matveev, Yue and Glazman\cite{glazman} used a
renormalization group argument to sum the log divergent corrections to all orders,
to obtain the exact transmission $\mathcal{T}(E)$.

For non interacting electrons, the QSHI point contact is characterized by a
$4\times 4$ scattering matrix $S_{ij}$ which relates the incoming wave in lead $i$ to
the outgoing wave in lead $j$,
\begin{equation}
\label{psiinout}
|\psi_{i,\rm out}\rangle = S_{ij} |\psi_{j,\rm in}\rangle.
\end{equation}
In terms of $S_{ij}$ the four terminal conductance is
\begin{equation}
\label{gijsij}
G_{ij} = {e^2\over h}(\delta_{ij} - |S_{ij}|^2).
\end{equation}

Under time reversal
$\Theta |\psi_{i,\rm out(in)}\rangle = +(-) Q_{ij} |\psi_{j,\rm in(out)}\rangle$, where
$Q = {\rm diag}(1,-1,1,-1)$.  This leads to the constraint
$S = - Q S^T Q$.  This combined with unitarity $S^\dagger S = 1$
allows $S$ to be parameterized as
\begin{equation}
\label{smatrix}
S = U^\dagger \left(\begin{array}{cccc}
0    &  t   & f    & r  \\
 t   &  0   & r^*  & -f^* \\
-f   &  r^* & 0    & -t ^* \\
 r   &  f^* & -t^* & 0
\end{array}\right) U,
\end{equation}
where $U_{ij} = \delta_{ij} e^{i\chi_i}$ is an unimportant gauge
transformation.  The complex numbers $t$ and $r$ describe the amplitudes for
spin conserving transmission and reflection across the point contact, while $f$
describes the amplitude for tunneling across the junction, combined with a spin
flip.  $f=0$ if spin is conserved.  The conductance can be expressed in terms
of the transmission probabilities
$\mathcal{R}= |r|^2$, $\mathcal{T} = |t|^2$ and $\mathcal{F} = |f|^2$, which
satisfy $\mathcal{R}+\mathcal{T}+\mathcal{F}=1$.
We find
\begin{eqnarray}
\label{gab(S)}
G_{XX} &=& {2 e^2\over h}(\mathcal{T}+\mathcal{F}) \nonumber \\
G_{YY} &=& {2 e^2\over h}(\mathcal{R}+\mathcal{F}) \\
G_{ZZ} &=& {2 e^2\over h}(1-\mathcal{F})\nonumber\\
G_{AB} &=& 0 \quad {\rm for}\ A\ne B.\nonumber
\end{eqnarray}
For a generic four terminal conductance device time reversal symmetry
guarantees only the reciprocity relation\cite{reciprocity} $G_{ij} = G_{ji}$, (or equivalently
$G_{AB}=G_{BA}$).  For the QSHI point
contact, the spin filtered nature of the edge states leads to additional
constraints.  First, the amplitude for an electron to be reflected back into
the lead it came from is $S_{ii} = 0$.  Thus $G_{ii} = e^2/h$.  A second less
obvious constraint is that $G_{13} = G_{24}$, which when combined with
reciprocity and unitarity is equivalent to $G_{12} = G_{34}$ and $G_{14} =
G_{23}$.  This leads to the vanishing of the skew conductance $G_{XY}$ as well
as $G_{XZ}$ and $G_{YZ}$
even when mirror symmetries ${\cal M}_X$ and ${\cal M}_Y$ are explicitly
violated.  This is a property of the non interacting electron model and can be
violated with electron electron interactions if the mirror symmetries are
absent.

In order to compute the renormalization of the $S$ matrix due to interactions
it is useful to study the perturbative expansion of the single electron thermal Green's
function, which can be represented as a matrix in the lead indices $i$, $j$ as
well as the channel labels $a={\rm in}/{\rm out}$.
\begin{equation}
\label{greensfunction}
{\mathbb{ G}}^{ab}_{ij}(x,\tau;x',\tau') = -i \langle T_\tau[ \psi_{i,a}(x,\tau)
\psi^\dagger_{j,b}(x',\tau')]\rangle,
\end{equation}
where $T_\tau$ denotes imaginary time ordering.
For non interacting electrons we have
\begin{equation}
\label{g0}
{\mathbb G}_{ij}(z,z') = {1\over{2\pi i}}\left(
\begin{Large}\begin{array}{cc}
{\delta_{ij}\over{z-z'}} &
 {S^*_{ji} \over{z - \bar z'}} \\
{S_{ij} \over {\bar z - z'}} &
{\delta_{ij}\over {\bar z - \bar z'}}
\end{array}
\end{Large}\right).
\end{equation}
where $z = \tau + ix$ and $\bar z= \tau-ix$, and the $a=$in/out
indices are displayed in matrix form.

\begin{figure}
\centerline{ \epsfig{figure=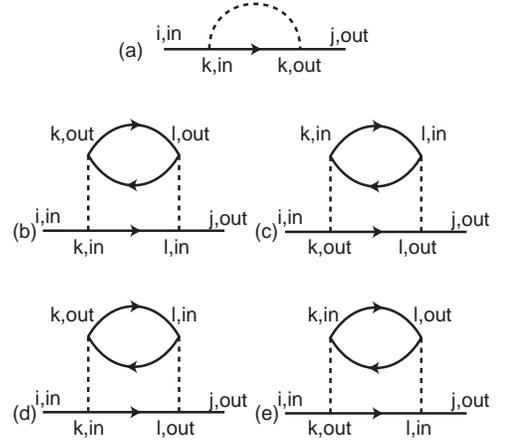,width=2.5in} }
 \caption{Feynman diagrams for the single electron Green's function.
 The dashed line is the interaction
 $u_2 (\psi_{\rm in}^\dagger \psi_{\rm in})(\psi_{\rm out}^\dagger \psi_{\rm out})$.
 The exchange diagram (a) vanishes because it involves $S_{kk}$, and diagrams
 (b) and (c) cancel one another.  (d) and (e) lead to a logarithmic
 correction to the S matrix given in (\ref{news}).
  }
 \label{diagrams}
\end{figure}

\begin{figure}
\centerline{ \epsfig{figure=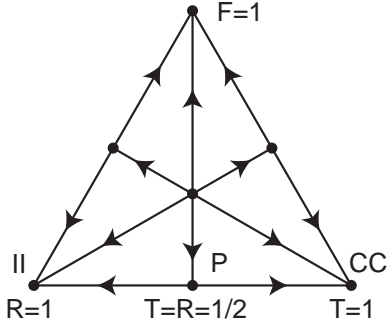,width=2in} }
 \caption{Renormalization group flow diagram for the transmission
 probabilities $\mathcal{T}$, $\mathcal{R}$ and $\mathcal{F}$ based on
 (\ref{dTRFdl}) represented in
 a ternary plot.  The CC, II and P fixed points of interest in this paper, which
 have $\mathcal{F}=0$ are on the bottom of the triangle.   }
 \label{triangle}
\end{figure}

We now compute the perturbative corrections to ${\mathbb G}^{\rm out,in}_{ij}$ using
the standard diagrammatic technique.  For simplicity, we adopt a model in which
$u_4=0$, so that the only interaction term involves
$u_2 (\psi_{\rm in}^\dagger \psi_{\rm in}) (\psi_{\rm out}^\dagger \psi_{\rm
out})$.  This considerably simplifies the analysis because many of the diagrams
are zero.  For instance, the exchange diagram shown in Fig. \ref{diagrams}(a), which was responsible for the
renormalization in the single channel Luttinger liquid problem is zero because
it must involve $\mathbb{G}^{\rm in,out}_{kk}$.  This off diagonal
Green's function depends on $S_{kk}$ which is zero due to the time reversal
symmetry constraint.  From (\ref{g}),
$g=\sqrt{(2\pi v_F-\lambda_2)/(2\pi v_F+\lambda_2)} \sim 1 - \lambda_2/(2\pi v_F)$.
Thus for $g=1-\epsilon$ we may replace $u_2$ by $2\pi v_F \epsilon$.
The nonzero diagrams at second order in $u_2$ are
shown in Fig. \ref{diagrams}(b-e).  Evaluating the second order
diagrams gives a Green's function of the form
\begin{equation}
\label{newg}
\mathbb{G}^{\rm out,in}_{ij} = {1\over{2\pi i}} {S'_{ij} \over {\bar z - z'}}
\end{equation}
with
\begin{equation}
\label{news}
S'_{ij} = S_{ij} + {\epsilon^2\over 4}\log{\Lambda\over E}\left[
S_{ij}S_{ji}S^*_{ji} - \sum_{kl} S_{ik} S_{kl}S^*_{lk}S^*_{kl} S_{lj} \right],
\end{equation}
where $\Lambda$ are $E$ are ultraviolet and infrared cutoffs respectively.
The first term in the brackets was due to the diagram in Fig.
\ref{diagrams}(d), while the second term was from Fig.
\ref{diagrams}(e).  Diagrams \ref{diagrams}(b) and \ref{diagrams}(c)
cancelled each other.
Rescaling the cutoff $\Lambda \rightarrow \Lambda e^{-\ell}$ leads to a
renormalization group flow equation for $S_{ij}$,
\begin{equation}
\label{dsdl}
{dS_{ij}\over d\ell} = {\epsilon^2\over 4}\left[
S_{ij}S_{ji}S^*_{ji} - \sum_{kl} S_{ik} S_{kl}S^*_{lk}S^*_{kl} S_{lj} \right].
\end{equation}
It is useful to rewrite this in terms of the transmission probabilities
$\mathcal{T},\mathcal{R},\mathcal{F}$.
The renormalization group flow equation then can be written in the form
\begin{eqnarray}
\label{dTRFdl}
d\mathcal{T}/d\ell &=&
{\epsilon^2}\mathcal{T}(\mathcal{T}-\mathcal{T}^2-\mathcal{R}^2-\mathcal{F}^2) \nonumber\\
d\mathcal{R}/d\ell &=&
{\epsilon^2}\mathcal{R}(\mathcal{R}-\mathcal{T}^2-\mathcal{R}^2-\mathcal{F}^2) \\
d\mathcal{F}/d\ell &=&
{\epsilon^2}\mathcal{F}(\mathcal{F}-\mathcal{T}^2-\mathcal{R}^2-\mathcal{F}^2).\nonumber
\end{eqnarray}
The flow diagram as a function of $\mathcal{R}$, $\mathcal{T}$ and
$\mathcal{F}$ is shown in Fig. \ref{triangle}.  There
are seven fixed points.  The bottom corners of the triangle are the stable fixed points
at $\mathcal{R}=1$, $\mathcal{T}=\mathcal{F}=0$ (the II phase)
and $\mathcal{T}=1$, $\mathcal{R}=\mathcal{F}=0$ (the CC phase).
The third stable fixed point at the top of the triangle with
$\mathcal{F}=1$, $\mathcal{T}=\mathcal{R}=0$, corresponds to the case where an incident electron is
transmitted perfectly with a spin flip.  This is presumably difficult to access
physically.  On the edges of the triangle are unstable fixed points describing
transitions between the different stable phases.  The critical fixed point P of interest
in this paper is the one on the bottom of the triangle at
$\mathcal{R}=\mathcal{T}=1/2$, $\mathcal{F}=0$.  Note that at
this fixed point the spin non conserving spin orbit processes, represented by $\mathcal{F}$ are
{\it irrelevant}.  At
the center of the triangle, at $\mathcal{R}=\mathcal{T}=\mathcal{F}=1/3$ is an unstable fixed point
describing a multicritical point.

To describe the critical fixed point P and the crossover to the II and CC phases
we now specialize to $\mathcal{F}=0$ and consider the flow equation for the single
parameter $\mathcal{T}$ characterizing the point contact,
\begin{equation}
\label{dTdl}
d\mathcal{T}/d\ell = -\epsilon^2 \mathcal{T}(1-\mathcal{T})(1-2\mathcal{T}).
\end{equation}
Eq. \ref{dTdl} can be integrated to determine the crossover scaling function.  If
at $\ell=0$ $\mathcal{T}=\mathcal{T}^0$, then,
\begin{equation}
\label{T(ell)}
\mathcal{T}(\ell) = {1\over 2}\left[1 + {\mathcal{T}^0 - 1/2 \over
\sqrt{(\mathcal{T}^0 - 1/2)^2 + \mathcal{T}^0(1-\mathcal{T}^0)
e^{-\epsilon^2 \ell}}}\right].
\end{equation}
As the gate voltage $V_G$ is adjusted through the pinch-off transition,
$\mathcal{T}^0$ passes
through 1/2 at $V_G = V_G^*$, so $\mathcal{T}^0- 1/2 \propto \Delta V_G$.   At
temperature $T$ we cut off the renormalization group flow at $\Lambda e^{-\ell}
\propto T$.  The conductance is then given by $G_{XX} = 2(e^2/h) \mathcal{T}(\ell =
\log(\Lambda/T))$.  For $\Delta V_G,T \rightarrow 0$ we define
$X = (2\mathcal{T}^0-1) e^{\epsilon^2\ell/2} \propto \Delta
V_G/T^{\epsilon^2/2}$ and write the conductance in the scaling form,
\begin{equation}
G_{XX}(\Delta V_G,T) = 2 {e^2\over h} {\cal G}_{1-\epsilon}(c{\Delta V_G\over
T^{\alpha_g}}),
\end{equation}
where $c$ is a non universal constant, the critical exponent is
\begin{equation}
\alpha_{1-\epsilon} = \epsilon^2/2,
\end{equation}
and
\begin{equation}
{\cal G}_{1-\epsilon}(X) = {1\over 2}\left[1 + {X \over\sqrt{1+X^2}}\right].
\end{equation}

We find that the logarithmic renormalization to the S matrix accounts for the
{\it only} correction to the conductance to linear order in
$\epsilon$.  In principle one must consider a ``RPA like" diagram for
the conductance evaluated by the Kubo formula.  While this gives a
correction for an {\it infinite} Luttinger liquid at finite
frequency, the correction is zero for a finite Luttinger liquid
connected to leads in the $\omega\rightarrow 0$ limit\cite{maslov, safi, ponomarenko,
kawabata}.  Since the critical conductance satisfies $G_g^*= 1- G_{1/g}^*$
it follows that $G_{1-\epsilon}^* = 1/2 + O(\epsilon^3)$.

\subsection{$g = 1/2 + \epsilon$}

$g=1/2$ is at the boundary where the CC and II phases become unstable and the
IC phase becomes stable.  We will show that when $g=1/2 + \epsilon$ the
critical fixed point describing the transition between the CC and II phases
approaches the IC fixed point and can be accessed perturbatively using theory
developed in Section II.B.3.  In addition, when $g=1/2$, the marginal operators $v_\rho
\cos 2\theta_\rho $ at the CC fixed point and $\tilde t
\cos\tilde\theta_\rho$ at the IC fixed point
can be expressed in terms of fictitious fermion
operators.  This fermionization process allows the entire crossover between the
CC and IC phases to be described using a non interacting fermion Hamiltonian.
A similar fermionization procedure can be used to describe the crossover
between the II and IC phases, which connect the marginal operators
$\tilde v_\sigma \cos\theta_\sigma$ and $\tilde t_\sigma \cos 2\tilde
\theta_\sigma$.
This will allow us to compute the full crossover scaling function ${\cal
G}_g(X)$ for $g=1/2+\epsilon$.

We will begin by discussing the perturbative analysis of the IC fixed point and
then go on to describe the fermionization procedure.

\subsubsection{Perturbative Analysis}

The IC fixed point is described by (\ref{s0ic}, \ref{s1ic}).  When $g=1/2+\epsilon$ the perturbations
$\tilde t_\rho \tau^x \cos \tilde \theta_\rho$ and $\tilde t_\sigma \tau^z
\cos\theta_\sigma$ both have scaling dimension $\Delta = 1 - 2\epsilon$, so the
IC fixed point is weakly unstable.  When $\tilde v_\sigma = 0$, nonzero $\tilde
t_\rho$ is expected to drive the system to the CC phase, while for $\tilde
t_\rho = 0$ nonzero $\tilde v_\sigma$ will drive the system to the II phase.
Thus, when both $\tilde t_\rho$ and $\tilde v_\sigma$ are non zero there must
be an unstable fixed point which separates the two alternatives.  This fixed
point can be described by considering the renormalization group flow equations
to {\it third} order in $\tilde v_\sigma$ and $\tilde t_\rho$.

\begin{figure}
\centerline{ \epsfig{figure=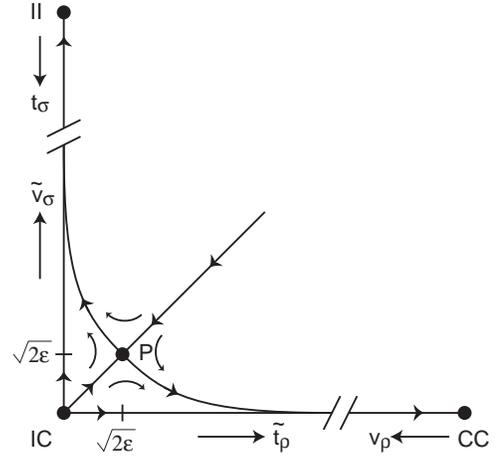,width=2.5in} }
 \caption{Renormalization group flow diagram characterizing the critical fixed point
 P for $g=1/2+\epsilon$.  When $\tilde v_\sigma$ and $\tilde t_\rho$ are small, the
 flows are given by (\ref{dtildetdl}).  On the axis $\tilde v_\sigma=0$ the fermionization
 procedure outlined in section III.D.2 determines the entire crossover between
 the IC and CC fixed points.  A similar theory describes the crossover between
 the IC and II fixed points for $\tilde t_\rho=0$.}
 \label{rgflow}
\end{figure}

The first order renormalization group equation for $\tilde t_\rho$ is
determined by the scaling dimension $\Delta(\tilde t_\rho)$.
The next nonzero term occurs at order $t_\rho
v_\sigma^2$.  To compute this term it is sufficient to use the theory at
$\epsilon = 0$.  Consider the third order term in the cumulant expansion
of the partition function, when fast degrees of freedom integrated out:
\begin{eqnarray}
\label{cumulant}
{1\over 2}\int d\tau_1 d\tau_2 \{\langle T_\tau[O_\rho(\tau)
O_\sigma(\tau_1) O_\sigma(\tau_2)]\rangle_> \nonumber \\
- \langle O_\rho(\tau) \rangle_>
\langle T_\tau[O_\sigma(\tau_1) O_\sigma(\tau_2)]\rangle_>\}.
\end{eqnarray}
Here $O_\rho = (\tilde t_\rho/\tau_c) \tau^x \cos\tilde\theta_\rho$ and
$O_\sigma = (\tilde v_\sigma/\tau_c) \tau^z \cos\theta_\sigma$.  $T_\tau$ indicates time
ordering, and $\langle \cdot \rangle_>$ denotes a trace over degrees of
freedom with $\Lambda/b < \omega < \Lambda$, and we assume for simplicity $b\gg 1$.
Since $\tilde\theta_\rho$ and $\theta_\sigma$ are independent and
commute with one another the other disconnected terms all cancel.
Moreover, the two terms in (\ref{cumulant}) will cancel each other unless
the time ordering of the $\tau^x$ and $\tau^z$ operators leads to a
relative minus sign between them,
\begin{eqnarray}
\label{timeorder}
 &&\langle T_\tau[ O_\rho(\tau)
O_\sigma(\tau_1) O_\sigma(\tau_2)]\rangle_> =\nonumber \\
&&\quad\quad\quad s_\pm \langle O_\rho(\tau) \rangle_>
\langle T_\tau[O_\sigma(\tau_1) O_\sigma(\tau_2)]\rangle_>,
\end{eqnarray}
where $s_\pm = {\rm sgn}(\tau-\tau_1)(\tau-\tau_2)$.  Thus the
pseudospin operators in (\ref{s1ic}) play a crucial role in the renormalization
of $\tilde t_\rho$.  Using the fact that
$\langle T_\tau[O_\sigma(\tau_1) O_\sigma(\tau_2)]\rangle_> = \tilde
v_\sigma^2/2(\tau_1-\tau_2)^2$ for $\epsilon=0$ we find that
the third order correction to $\tilde t_\rho$ is
$\delta \tilde t_\rho = - t_\rho v_\sigma^2 \log b$.
This leads to the renormalization group flow equation for $\tilde
t_\rho$, along with a corresponding equation for $\tilde v_\sigma$,
\begin{eqnarray}
\label{dtildetdl}
d \tilde t_\rho/d\ell &=& 2\epsilon \tilde t_\rho -  \tilde t_\rho \tilde
v_\sigma^2 \nonumber\\
d \tilde v_\sigma/d\ell &=& 2\epsilon \tilde v_\sigma -  \tilde v_\sigma \tilde
t_\rho^2.
\end{eqnarray}
The renormalization group flow diagram is shown in Fig. \ref{rgflow}.  There is an
unstable fixed point P at $\tilde t_\rho = \tilde v_\sigma = \sqrt{2\epsilon}$,
with a single relevant operator.
P separates the flows to the CC and II phases
for which which $\tilde t_\rho$ or $\tilde v_\sigma$
grow.  Note that spin orbit terms such as $v_{so}$ and $v_{sf}$ discussed
in Section II.B.5 are irrelevant at P (see
Eq. \ref{vso}).  This perturbative calculation
provides further evidence that P exhibits
emergent spin conservation, as well as emergent mirror symmetry.
The critical exponent associate with the single relevant operator a P
is
\begin{equation}
\label{alpha1/2+ep}
\alpha_{1/2+\epsilon} = 4\epsilon.
\end{equation}

The Kubo conductance $G^K_{\rho\rho}$
at the fixed point can be computed from (\ref{kubo}) by identifying the
current operator
\begin{equation}
\label{irho}
I_\rho = (\tilde t_\rho/\tau_c) \sin \tilde\theta_\rho.
\end{equation}
This leads to
\begin{equation}
\label{gkrhorho}
G^K_{\rho\rho} = {e^2\over h}\pi^2 \tilde t_\rho^2.
\end{equation}
It is useful to define
$\mathcal{T}_\rho = \pi^2 \tilde t_\rho^2$.  We will see in the following
section that this can be interpreted as a transmission probability
for fictitious free fermions that describe the problem at $g=1/2$.
In terms of $\mathcal{T}_\rho$ (noting that $\mathcal{T}_\rho\ll 1$ in this perturbative
regime) we may use (\ref{gxxgkubo}) to write the physical
conductance as
\begin{equation}
\label{gxxt1/2}
G_{XX} = {e^2\over h} \mathcal{T}_\rho.
\end{equation}
A similar calculation gives
\begin{equation}
\label{gyyr1/2}
G_{YY} = {e^2\over h} \mathcal{R}_\sigma,
\end{equation}
where $\mathcal{R}_\sigma = \pi^2 \tilde v_\sigma^2$ can similarly be
interpreted as a reflection probability for a different fictitious free fermion
at $g=1/2$.
At the critical fixed point
$\mathcal{T}_\rho=\mathcal{R}_\sigma=2\pi^2\epsilon$.  Thus,
\begin{eqnarray}
G^*_{XX} &=& G^*_{YY} =2  {e^2\over h} \pi^2 \epsilon. \nonumber \\
G^*_{XY} &=& 0.
\end{eqnarray}

The behavior away from the critical point can be determined by integrating (\ref{dtildetdl}).
To this end it is helpful to rewrite (\ref{dtildetdl}) in terms of $\mathcal{T}_\rho$ and
$\mathcal{R}_\sigma$ in the form
\begin{eqnarray}
\label{dTRdl}
d(\mathcal{T}_\rho-\mathcal{R}_\sigma)/d\ell &=&
4\epsilon(\mathcal{T}_\rho-\mathcal{R}_\sigma) \nonumber\\
d\log(\mathcal{T}_\rho/\mathcal{R}_\sigma)/d\ell &=& (2/{\pi^2})
(\mathcal{T}_\rho-\mathcal{R}_\sigma).
\end{eqnarray}
%
If $(\mathcal{T}_\rho,\mathcal{R}_\sigma) =
(\mathcal{T}_\rho^0,\mathcal{R}_\sigma ^0)$ for $\ell = 0$, then we find
\begin{eqnarray}
\label{TR(l)}
\mathcal{T}_\rho(\ell) &=& {(\mathcal{T}^0_\rho - \mathcal{R}^0_\sigma)
e^{4\epsilon \ell} \over {1- {\mathcal{R}^0_\sigma\over {\mathcal{T}^0_\rho}}
\exp\left[-{\mathcal{T}^0_\rho-\mathcal{R}^0_\sigma\over
{2\pi^2\epsilon}}(e^{4\epsilon \ell}-1)\right]}} \nonumber\\
\mathcal{R}_\sigma(\ell) &=& {(\mathcal{R}^0_\sigma-\mathcal{T}^0_\rho)
e^{4\epsilon \ell} \over {1-{\mathcal{T}^0_\rho\over{\mathcal{R}^0_\sigma}}
\exp\left[-{\mathcal{R}^0_\sigma-\mathcal{T}^0_\rho\over
{2\pi^2\epsilon}}(e^{4\epsilon \ell}-1)\right]}}.
\end{eqnarray}

At the pinch-off transition $V_G = V_G^*$, $\mathcal{R}_0=\mathcal{T}_0$.
Thus, $\mathcal{T}_0 - \mathcal{R}_0 \propto
\Delta V_G$.  At temperature $T$ we cut off the renormalization group flow at
$\Lambda e^{-\ell} \propto T$.  Thus, in the limit $\Delta V_G,T
\rightarrow 0$ we define $X =
(\mathcal{T}^0_\rho-\mathcal{R}^0_\sigma) e^{4\epsilon\ell}/2 \propto
\Delta V_G/T^{4\epsilon}$.  The conductance then has the form
\begin{eqnarray}
G_{XX}(\Delta V_G,T) &=& 2 {e^2\over h} {\cal G}_g\left(c {\Delta
V_G\over {T^{\alpha_g}}}\right) \nonumber\\
G_{YY}(\Delta V_G,T) &=& 2 {e^2\over h} {\cal G}_g\left(-c {\Delta
V_G\over {T^{\alpha_g}}}\right),
\end{eqnarray}
with
\begin{equation}
\label{pscaling1/2+ep}
{\cal G}_{1/2+\epsilon}(X) = {X \over{1-e^{-X/(\pi^2\epsilon)}}}.
\end{equation}

This perturbative calculation is only valid when $\mathcal{T}_\rho,\mathcal{R}_\sigma\ll 1$.
Thus (\ref{pscaling1/2+ep}) breaks
down at low temperature, since as the energy is lowered either $\mathcal{T}_\rho$ or
$\mathcal{R}_\sigma$ grows.  (\ref{pscaling1/2+ep}) is valid as long as $|X| \ll 1$.
Note, however that when $\epsilon
\ll 1$ we have ${\cal G}_{1/2+\epsilon}(X) = X \theta(X)$ when $\epsilon \ll X
\ll 1$.  In this regime, the smaller of $\mathcal{T}$ and $\mathcal{R}$ has gone to zero.  Thus
we have
\begin{eqnarray}
\label{TRasymptote}
\left.\begin{array}{l}
\mathcal{T}_\rho(\ell) = (\mathcal{T}^0_\rho - \mathcal{R}^0_\sigma) e^{4\epsilon\ell}\\
\mathcal{R}_\sigma(\ell) = 0
\end{array}
\right\}  &  \ (\mathcal{T}^0_\rho - \mathcal{R}^0_\sigma) > 0 \nonumber \\
\left.\begin{array}{l}
\mathcal{T}_\rho(\ell) = 0\\
\mathcal{R}_\sigma(\ell) = (\mathcal{R}^0_\sigma -\mathcal{T}^0_\rho )
e^{4\epsilon\ell}
\end{array} \right\} &  \ (\mathcal{R}^0_\sigma -\mathcal{T}^0_\rho )>
0,
\end{eqnarray}
 and the
unstable flow is either on the x or y axis of Fig. \ref{rgflow}.
In the next section we will solve the crossover exactly on these lines.  This
will allow us to compute the ${\cal G}_{1/2+\epsilon}(X)$ exactly for all $X$.

\subsubsection{Fermionization}

In this subsection we study the crossover between the IC fixed
point and the CC and II fixed points for $g=1/2+\epsilon$.  There are two cases
to consider.  First, for $\Delta V_G>0$ we will study the crossover between the IC and CC
on the horizontal axis of Fig. \ref{rgflow} with $\tilde v_\sigma=0$.
This problem can be mapped to a single channel one dimensional
fermi gas with weak electron electron interactions proportional to $\epsilon$.
This allows us to use the method of Matveev, Yue and Glazman\cite{glazman} to
compute the crossover scaling functions ${\cal G}_{g,XX}(X)$ and ${\cal
G}_{g,YY}(X)$ for $X>0$ exactly.
For $\Delta V_G<0$ the
crossover between the IC and II fixed points is on the vertical
axis of Fig. \ref{rgflow} with $\tilde t_\rho=0$.
This can be fermionized by introducing a {\it different} set of
free fermions to compute the scaling functions for $X<0$.  The latter
calculation (which is virtually identical to the former)
is unnecessary, however, because we can use (\ref{gsymmetry-x}) to deduce the
scaling functions for $X<0$.  We will therefore focus on the IC to CC
crossover.

The crossover between the IC and the CC fixed points can be described
by the action in the CC limit
\begin{equation}
\label{sccg=1/2}
S_{CC} = {1\over\beta}\sum_{n} {1\over {2\pi g}}|\omega_n||\theta_\rho(\omega_n)|^2 + \int
d\tau v_\rho \cos 2\theta_\rho.
\end{equation}
$v_\rho\ll 1$  describes the CC phase.  When $v_\rho\gg 1$ the dual theory,
formulated as in section II.B.3 in terms of instantons with amplitude $\tilde t_\rho$,
describes the IC phase.  When $\tilde v_\sigma=0$ at the IC fixed point we can
safely ignore the pseudospin, and set $\tau^x=1$.

For $g=1/2$ this model is equivalent to the bosonized representation of a
weak link in a single channel {\it non interacting} fermion with weak backscattering.
\begin{equation}
\label{hfermion}
{\cal H}_f = -i v \tilde\psi^\dagger \partial_x\sigma^z\tilde \psi
 + v_f \tilde\psi^\dagger \sigma^x
\tilde\psi \delta(x).
\end{equation}
where $\tilde\psi = (\tilde\psi_R,\tilde\psi_L)^T$
is a two component fermion operator describing
right and left movers.  Using the bosonization relation (\ref{bosonizepsi}) we
identify $2\theta_\rho = \phi_R-\phi_L$ and $v_f = \pi
v_\rho/v$.
The free fermion problem is solvable and characterized by a
transmission probability $\mathcal{T}_\rho = {\rm sech}^2(v_f/v)$.
The free fermion solution therefore connects the CC limit
($\mathcal{T}_\rho=1$) with the IC limit ($\mathcal{T}_\rho=0$).

The Kubo conductance $G^K_{\rho\rho}$ may be computed with the
identification $J_\rho = \partial_t \theta_\rho/\pi = v \tilde\psi^\dagger
\sigma^z \tilde\psi$, giving
\begin{equation}
G^K_{\rho\rho} = {e^2\over h}\mathcal{T}_\rho.
\end{equation}
Note that this is the same as (\ref{gxxt1/2}), derived in the opposite limit near
the IC fixed point.  When $v_\rho$ is large,
$\mathcal{T}_\rho \ll 1$, and we can
identify $\mathcal{T}_\rho = (\pi \tilde t_\rho)^2$.
The physical conductance, measured with leads can be determined
following the analysis in appendix A.  From (\ref{gxxgkubo}) we find
\begin{equation}
\label{gxxtrho}
G_{XX} = 2 {e^2\over h} {\mathcal{T}_\rho\over{2-\mathcal{T}_\rho}}.
\end{equation}
Since $v_\sigma=0$ in (\ref{sccg=1/2}), we have
\begin{equation}
\label{gyytrho}
G_{YY} = 0.
\end{equation}

For $g=1/2+\epsilon$ the IC fixed point becomes slightly unstable,
while the CC fixed point becomes slightly stable.  In this case
the free fermion problem includes a weak {\it
attractive} interaction
\begin{equation}
{\cal H}_f^{\rm int} = - u_f (\tilde\psi_L^\dagger
\tilde\psi_L)(\tilde\psi_R^\dagger\tilde\psi_R),
\end{equation}
with $u_f = 2\pi v \epsilon$.  This leads to a logarithmic
renormalization of $\mathcal{T}_\rho$, which drives a crossover to the
CC limit.  The correction to $\mathcal{T}_\rho$ occurs at {\it first} order
in $u_f$, and is due to the exchange diagram, shown in Fig. \ref{diagrams}(a).
The analysis is exactly the same as that performed by
Matveev, Yue and Glazman.  As in section III.C the result can be cast in terms of a
renormalization group flow equation for ${\cal T}_\rho$.
\begin{equation}
\label{dTdlT(1-T)}
d \mathcal{T_\rho}/d\ell = 4 \epsilon \mathcal{T}_\rho(1-\mathcal{T}_\rho).
\end{equation}
Integrating (\ref{dTdlT(1-T)}) gives
\begin{equation}
\label{T(l)fermionized}
\mathcal{T}_\rho(\ell) = {\mathcal{T}_\rho^0 e^{4\epsilon\ell}\over{1+
\mathcal{T}_\rho^0( e^{4\epsilon\ell}-1)}},
\end{equation}
where $\mathcal{T}_\rho^0 = \mathcal{T}_\rho(\ell=0)$.
The scaling function for $\Delta V_G >0$ then follows by using
the initial condition from (\ref{TRasymptote}), so that
$\mathcal{T}^0_\rho \propto \Delta V_G$.
Then, for $\Delta V_G,T \rightarrow 0$ we define
$X = \mathcal{T}_\rho^0 e^{4\epsilon
\ell}/2 \propto \Delta V_G/T^{4\epsilon}$.  Using (\ref{gxxtrho}), (\ref{gyytrho}) and
(\ref{T(l)fermionized}),
the conductance has the scaling form for $X>0$
\begin{eqnarray}
{\cal G}_{XX, 1/2+\epsilon}(X) &=&  {X\over{X+1}} \nonumber \\
{\cal G}_{YY, 1/2+\epsilon}(X) &=& 0.
\end{eqnarray}
Using (\ref{gsymmetry-x}), we may deduce the corresponding behavior for $\Delta
V_G<0$ (or $X<0$).  The scaling function then has the form
\begin{equation}
\label{fscaling1/2+ep}
{\cal G}_{1/2+\epsilon}(X) =  \theta(X){X\over{X+1}}.
\end{equation}
Note that for $X\ll 1$ ${\cal G}_{1/2+\epsilon}(X) = X\theta(X)$, in
agreement with the limiting behavior of (\ref{pscaling1/2+ep}) for $X \gg \epsilon$.  These
two expressions can thus be combined to give
\begin{equation}
{\cal G}_{1/2+\epsilon}(X) =  {X\over{X+1-e^{-X/(\pi^2\epsilon)}}},
\end{equation}
which reproduces (\ref{pscaling1/2+ep}) when $|X| \sim \epsilon \ll 1$ and
(\ref{fscaling1/2+ep}) when $|X| \gg \epsilon$.  This function is plotted
in Fig. \ref{scalingfig}(b).  Note, however, that
this formula does not correctly capture the leading behavior for
$X<0$ when $|X| \gg \epsilon$.  In particular, it misses the
$X\rightarrow -\infty$ behavior, which (\ref{gasymptote}) and (\ref{beta})
predict is proportional to $|X|^{-1/(8\epsilon)}$.  This regime is
analyzed in the following section.

\subsubsection{Rebosonization}

We now analyze the leading behavior of ${\cal G}_{1/2+\epsilon}(X)$ for
$X<0$ and $|X|\gg\epsilon$ when $\epsilon$ is small.  Equivalently, we consider
${\cal G}_{1/2+\epsilon,YY}(X)$ for $X>0$.  This requires
extending the renormalization group flow equation for  $\tilde v_\sigma$
 given in (\ref{dtildetdl})
to all $\tilde t_\rho$ (or equivalently $\mathcal{T}_\rho$).  This can be
done by using the fermionized representation of
$\tilde t_\rho\tau^x \cos\tilde\theta_\rho$ in (\ref{s1ic}).
The key point is that the presence of the pseudospin operator $\tau^x$
means that the operator $\tilde v_\sigma \tau^z
\cos\theta_\sigma$ changes the sign of the transmission amplitude for
the fermions $\tilde\psi$.  This results in an X ray edge like
contribution to the renormalization of $\tilde v_\sigma$.  This can
be computed by a method analogous to that used by Schotte and
Schotte\cite{schotte}
to solve the X ray edge problem, which involves transforming the
non interacting fermions to even
and odd parity scattering states and then rebosonizing.  This
approach was used to study the X ray edge problem in a Luttinger
liquid in Ref. \onlinecite{kaneglazman}.

We begin by writing (\ref{s1ic}),
${\cal H} = {\cal H}_\sigma + {\cal H}_\rho$ with
\begin{equation}
{\cal H}_\sigma = {\cal H}_\sigma^0 + \tilde v_\sigma \tau^z
\cos\theta_\sigma
\label{fermion1}
\end{equation}
and
\begin{equation}
{\cal H}_\rho = -i v \tilde\psi^\dagger \sigma^z\partial_x \tilde
\psi + t_f \tau^x \tilde \psi^\dagger \sigma^x \tilde \psi \delta(x).
\label{fermion2}
\end{equation}
Here ${\cal H}^0_\sigma$ is the $\sigma$ part of (\ref{h0thetaphi}), and
we explicitly account for the pseudospin $\tau^x$.
Eq. \ref{fermion2} can be rebosonized by first replacing $\psi_2(x)
\rightarrow \psi_2(-x)$, which transforms the non chiral fermions
to chiral fermions, eliminating the $\sigma^z$ in the first term,
but leaving the second term alone.  Then we perform a SU(2) rotation
$(\tilde\psi_1,\tilde\psi_2)\rightarrow (\tilde\psi_e,\tilde\psi_o)$,
which changes $\sigma^x$ in the
second term into $\sigma^z$.  $\tilde\psi_{e(o)}$ describe the even (odd) parity
scattering states characterized by scattering phase shifts
$\delta_e=-\delta_o$ that specify $\tilde\psi_{e(o)}(x>0) =
e^{2 i\delta_{e(o)}}\tilde\psi_{e(o)}(x<0)$.  We next bosonize $\tilde\psi_{e,o}
\rightarrow e^{i\phi_{e,o}}/\sqrt{2\pi x_c}$ and define $\phi_\pm =
\phi_e \pm \phi_o$.  Then
\begin{equation}
{\cal H}_\rho = {v\over{8\pi}}\left[(\partial_x\phi_+)^2 +
(\partial_x\phi_-)^2\right] + {v\over{2\pi}}\delta_- \tau^x (\partial_x \phi_-) \delta(x)
,
\label{hrho2}
\end{equation}
where $\phi_\pm$ obey,
$[\phi_\pm(x),\phi_\pm(x')] = 2\pi i{\rm sgn}(x-x')$.
$\delta_-=\delta_e-\delta_o$
is related to the transmission probability by
\begin{equation}
\mathcal{T}_\rho = \sin^2\delta_-.
\label{sindeltarho}
\end{equation}

$\delta_-$ can be eliminated from (\ref{hrho2}) by the canonical
transformation $U = \exp[i \tau^x  \delta_-\phi_-(x=0)/(2\pi)]$, which
shifts $\phi_- \rightarrow \phi_- + {\rm sgn}(x) \delta_- \tau^x$.
This transformation also rotates $\tau^z$ in
(\ref{fermion1}), which becomes
\begin{equation}
{\cal H}_\sigma = H^0_\sigma + \tilde v_\sigma\left[\tau^+
e^{i \phi_- \delta_- /\pi} +
\tau^-e^{-i \phi_- \delta_- /\pi}\right] \cos \theta_\sigma
\end{equation}
where $\tau^\pm = \tau^z \pm i\tau^y$.
The renormalization of $\tilde v_\sigma$ can then easily be
determined for arbitrary $\delta_-$.  We find
\begin{equation}
\label{dRdl}
{d\tilde v_\sigma\over {d\ell}} = \left( 2 \epsilon
- \left({\delta_-\over\pi}\right)^2\right)\tilde v_\sigma.
\end{equation}
For small $\tilde t_\rho$, $\delta_- = \pi \tilde t_\rho$, and
(\ref{dRdl}) reproduces (\ref{dtildetdl}).  However, (\ref{dRdl}) remains
valid to lowest order in $\epsilon$ for all $\mathcal{T}_\rho$.

We now integrate (\ref{dTRdl}) to a scale $\ell_0$
where from (\ref{TR(l)}) $\mathcal{T}_\rho(\ell_0) = 2 X_0$ and
$\mathcal{R}_\sigma(\ell_0) = 2X_0 e^{-X_0/(\pi^2\epsilon)}$
is small.  (Here $X_0 = (\mathcal{T}^0_\rho-\mathcal{R}^0_\sigma)
e^{4\epsilon\ell_0}/2$.)
We then use that as an
initial value for (\ref{dRdl}), which we integrate assuming
$\mathcal{T}_\rho(\ell)$ is given by (\ref{T(l)fermionized}) and
is unaffected by the small $\mathcal{R}_\sigma$.  Expressing
(\ref{T(l)fermionized}) in terms of (\ref{sindeltarho}) we have
\begin{equation}
\delta_-(\ell) = \rm tan^{-1} \left[\delta_-(\ell_0)
e^{2\epsilon(\ell-\ell_0)}\right]
\end{equation}
where $\delta_-(\ell_0) = \sin^{-1}\sqrt{\mathcal{T}_\rho(\ell_0)} \sim
\sqrt{2X_0}$.  As before, we define
$X= (\mathcal{T}^0_\rho-\mathcal{R}^0_\sigma)
e^{4\epsilon\ell}/2$.
We may express $G_{YY} =
(e^2/h)\mathcal{R}_\sigma$ with $\mathcal{R}_\sigma = \pi^2 \tilde
v_\sigma^2$.  Integrating (\ref{dRdl}) we then find
\begin{equation}
G_{YY}(X) = 2{e^2\over h}X e^{- F(X)/\epsilon},
\end{equation}
where
\begin{equation}
F(X) = {1\over\pi^2} \int_0^{\sqrt{2X}} {dx\over x}
\left(\tan^{-1} x \right)^2.
\end{equation}
Thus, for $X<0$, $|X|\gg\epsilon$  and $\epsilon\rightarrow 0$ we find
\begin{equation}
{\cal G}_{1/2+\epsilon}(X) = |X| e^{- F(|X|)/\epsilon}.
\end{equation}
The asymptotic behavior $F(X) = X/\pi^2$ for $|X|\ll 1$ reproduces
(\ref{pscaling1/2+ep}) when $|X|\gg\epsilon$.  For $|X|\gg 1$ we find
\begin{equation}
F(X\rightarrow \infty) = {1\over {8}} \log 2X -
{7\zeta(3)\over{4\pi^2}}.
\end{equation}
where $\zeta(3) = 1.20$ is the Riemann zeta function.  This gives the asymptotic behavior
\begin{equation}
{\cal G}_{1/2+\epsilon}(X\rightarrow -\infty) =
\left({e^{14\zeta(3)/\pi^2}\over{2|X|}}\right)^{1\over{8\epsilon}},
\end{equation}
which is quoted in Table \ref{asymptable}.

\section{Discussion and Conclusion}

In this paper we have examined several novel properties of a point contact in a
QSHI.  We showed that the pinch-off as a function of gate
voltage is governed by a non trivial quantum phase transition, which leads to
scaling behavior of the conductance as a function of temperature and gate
voltage characterized by a universal scaling function.  We computed this
scaling function and other properties of the critical point in certain solvable
limits which provide an overall picture of the behavior as a function of the
Luttinger liquid parameter $g$.

In addition, we showed that the four terminal conductance has a simple
structure when expressed in terms of the natural variables, $G_{AB}$,
and that at the low temperature fixed points, the leading corrections
to the different components of $G_{AB}$ can have different
temperature dependence.  In particular, we showed that the skew
conductance $G_{XY}$ vanishes as $T^\gamma$ with $\gamma\ge 2$.

Finally, we showed that for strong interactions, $g<1/2$, the stable
phase is the time reversal breaking insulating phase.  Transport in
that phase occurs via novel fractionalized excitations that have
clear signatures in noise correlations.

There are a number of problems for future research that our work raises.  We will
divide the discussion into experimental and theoretical issues.

\subsection{Experimental Issues}

The QSHI has been observed in transport experiments on
HgTe/HgCdTe quantum well structures.  A crucial issue is the value of
the interaction parameter $g$.  A simple estimate can be developed
based on the long range Coulomb interaction\cite{glazman2}.
First consider the limit $\xi \gg w$, where
$w$ is the quantum well width and $\xi$ is the
evanescent decay length of the edge state wavefunction into the bulk
QSHI.  We model the edge state as a two dimensional charged sheet with a charge
density profile proportional to $\theta(x) \exp(-2 x/\xi)$, a distance
$d$ above a conducting ground plane.  The long range interaction then leads to
$u_2 = u_4 = (2e^2/\epsilon)\log(4 e^\gamma d/\xi)$, where $\epsilon$
is the dielectric constant and $\gamma = .577$ is Euler's constant.
As a second model, assume $\xi \ll w$, and model the edge state as a uniformly charged two
dimensional strip of width $w$ perpendicular to a ground plane a distance $d$
away.  This gives $u_2 = u_4 = (2e^2/\epsilon)\log(2 e^{3/2} d/w)$.
The intermediate regime $\xi \sim w$ can be solved numerically, and we find
that it is accurately described by a simple interpolation between the above
limits with $4d/(\xi e^{-\gamma} + 2w e^{-3/2})$ in the log.  This leads to\cite{similar}.
\begin{equation}
g = \left[1+{2\over\pi} {e^2\over {\epsilon\hbar v_F}} \log \left(
{7.1 d\over{ \xi + 0.8 w}}\right)\right]^{-1/2}.
\end{equation}
For $\epsilon = 15$, $\hbar v_F = .35$eVnm, $\xi = 2\hbar v_F/E_{\rm gap} \sim 30$nm
($E_{\rm gap}$ is the gap of the bulk QSHI), $w=12$nm and $d =
150$nm\cite{review} this predicts $g\sim 0.8$.  The critical exponent governing the
temperature dependence of the pinch-off curve (\ref{scalingform}) is
then $\alpha_g \sim .02$.  In the CC and II phase the conductance vanishes as
$T^\delta$ with $\delta_g = g+g^{-1}-2 \sim .05$.

The good news is that since $g$ is close to $1$ the low temperature
scaling behavior should be accurately described by the scaling
function (\ref{g1-epsilon}) computed in the limit $g \rightarrow 1$.  The bad
news, is that the smallness of $\alpha_g$ and $\delta_g$ mean that it will be
difficult to see much dynamic range in the conductance as a function
of temperature.  Nonetheless, it may be possible to observe logarithmic
corrections to the conductance as a function of temperature, and
by comparing pinch-off curves at
different temperatures it may be possible to observe the predicted
sharpening of the transition as temperature is lowered.

The skew conductance $G_{XY}$ is predicted be zero for non interacting
electrons, and with weak interactions vanishes as $T^2$.  This is a
consequence of the unique edge state structure of the QSHI, and
remains robust when the interactions are weak.

To probe the critical behavior of the pinch-off transition, as well as
the more exotic strong interaction phases it would be desirable to
engineer structures with smaller $g$.  Perhaps this could be
accomplished by modifying either the dielectric environment or the
bare Fermi velocity of the edge states.  Maciejko et al. \cite{oreg}
have suggested that this may be possible using InAs/GaSb/AlSb type-II
quantum wells\cite{cooper,liu}.

\subsection{Theoretical Issues}

Our work points to a number of theoretical problems for future study.
It would be very interesting if the powerful framework of conformal
field theory can be used to analyze the intermediate critical fixed
point as well as the crossover scaling function.  Perhaps the first
place to look is $g=1/\sqrt{3}$.  Maybe it is possible to take
advantage of the triangular symmetry of the QBM
problem to develop a complete description of the critical fixed
point, analogous to the mapping to the 3 channel Kondo problem\cite{yikane} and
the 3 state Potts model\cite{affleck1} that apply in a different regime.
In the absence of an analytic solution, this problem is amenable to a
numerical Monte Carlo analysis analogous to the calculation of the
resonance crossover scaling function performed in Ref.
\onlinecite{moon}.

In addition, there are a number of other fixed points which we did
not analyze in detail in this paper.  (Recall for $g=1-\epsilon$ we
found seven).  It would be of interest to develop a more systematic
classification of all of the fixed points, analogous to the analysis of
three coupled Luttinger liquids performed by Oshikawa, Chamon and
Affleck and Hou
\cite{chamon2,chamon3}.

\acknowledgments

It is a pleasure to thank Claudio Chamon and Eun-Ah Kim for
introducing us to their work and Liang Fu for helpful discussions.
This work was supported by NSF grant DMR-0605066.

\begin{appendix}

\section{Four Terminal Conductance}

The electrical response of the point contact can be characterized by a four
terminal conductance,
\begin{equation}
I_i = \sum_j G_{ij} V_j,
\end{equation}
where $I_i$ is the current flowing into lead $i$ and $V_j$ is the voltage at
lead $j$.  In this appendix we will develop a convenient representation for
$G_{ij}$.  Section 1 shows that $G_{ij}$ can be characterized by a $3 \times 3$
matrix, whose entries have a clear physical meaning.  This
representation allows constraints due to symmetry to be expressed in a simple
way, which reduces the number of independent parameters characterizing the
conductance.  Finally, in section 3 we show how $G_{ij}$ is related to the
conductance of the SLL model computed by the Kubo formula.

\subsection{Conductance matrix}

The $4\times 4$ matrix $G_{ij}$ is constrained by current conservation to
satisfy $\sum_i G_{ij} = \sum_j G_{ij} = 0$.  In the absence of any symmetry
constraints, there are thus $9$ independent parameters characterizing $G_{ij}$.
In this section we will cast these $9$ numbers as a $3 \times 3$ matrix, in which each
of the entries has a clear physical meaning.  In this representation
constraints due to symmetry have a simple form.

Since the four currents $I_i$ satisfy $\sum_i I_i = 0$, they are
determined by three {\it independent} currents, which we define as
$I_A = (I_X,I_Y,I_Z)$, and satisfy
\begin{equation}
I_i = \sum_\alpha M_{i A} I_A,
\end{equation}
where the $4 \times 3$ matrix $M_{iA}$ is
\begin{equation}
\label{mmatrix}
M = {1\over 2}\left(\begin{array}{rrrr}
1 & 1 & 1  \\
-1 & 1 & -1 \\
-1 & -1 & 1 \\
1 & -1 & -1
\end{array}\right).
\end{equation}
$I_X=I_1+I_4$ is the total current flowing from left to right along the Hall bar,
whereas $I_Y=I_1+I_2$
is the current flowing from top to bottom.  The third current $I_Z=I_1+I_3$ is the
current flowing in on opposite leads (1 and 3) and flowing out in leads 2 and
4.  Similarly, the voltages $V_i$, which are defined up to an additive
constant, define three independent voltage differences
$V_\beta = (V_X,V_Y,V_Z)$, with
\begin{equation}
V_B = \sum_j M^T_{B j} V_j.
\end{equation}
$V_X$ biases leads 1 and 4 relative to leads 2 and 3, $V_Y$ biases
leads 1 and 2 relative to leads 3 and 4, and $V_Z$ biases leads 1 and
3 relative to leads 2 and 4.

The new currents and voltages are then related by a $3 \times 3$ conductance
matrix
\begin{equation}
\label{iavb}
I_A = \sum_\beta G_{AB} V_B.
\end{equation}
The 9 elements of $G_{AB}$ determine the four terminal
conductance matrix,
\begin{equation}
\label{gijgab}
G_{ij} = \sum_{AB} M_{iA} G_{AB} M^T_{B j}.
\end{equation}
The elements of $G_{AB}$ have a simple physical interpretation.
$G_{XX}$ is  the ``two terminal" conductance measured horizontally in Fig.
\ref{fig1}
by applying a voltage to leads 1 and 4 and measuring the current $I_1+I_4$.
Similarly $G_{YY}$ is a two terminal conductance measured vertically.  $G_{ZZ}$
describes a two terminal conductance defined by combining the opposite leads 1
and 3 together into a single lead (and similarly for leads 2 and 4).
$G_{XY}$
is a ``skew" conductance describing the current $I_1+I_4$ in response to
voltages applied to leads 1 and 2.  The other off diagonal conductances
can be understood similarly.

\subsection{Symmetry Constraints}

The form of $G_{AB}$ simplifies considerably in the presence of
symmetries.

\subsubsection{Time Reversal Symmetry}

In the presence of time reversal symmetry the four terminal conductance obeys
the reciprocity relation\cite{reciprocity}, $G_{ij} = G_{ji}$.  This implies $G_{AB} =
G_{BA}$.  Thus, with time reversal symmetry the conductance has 6
independent components.

\subsubsection{Spin Rotational Symmetry}

When the spin $S_z$ is conserved the current of up and down spins flowing into
the junction must independently be conserved.  It follows that
\begin{eqnarray}
I_{1,\rm in} + I_{3,\rm in} = I_{2,\rm out} + I_{4,\rm out} \nonumber\\
I_{2,\rm in} + I_{4,\rm in} = I_{1,\rm out} + I_{3,\rm out}.
\end{eqnarray}
Since in the Fermi liquid lead (where the interactions have been turned off) we
have $I_{i,\rm in} = (e^2/h)V_i$, this implies that
\begin{equation}
I_1 + I_3 = -I_2 - I_4 = {e^2\over h}(V_1 + V_3 - V_2 - V_4).
\end{equation}
It then follows that
\begin{eqnarray}
&G_{ZZ} = 2 e^2/h \nonumber\\
&G_{ZX} = G_{ZY} = 0.
\end{eqnarray}
Thus, which spin conservation the conductance is characterized by 3 components:
the two terminal conductances $G_{XX}$, $G_{YY}$ and the skew conductance
$G_{XY}$.

The quantization of $G_{ZZ}$ and vanishing of $G_{ZB}$ are therefore a
diagnostic for the conservation of spin.  Though spin orbit terms violating
$S_z$ conservation are generically present, we will argue that at the low
energy fixed points of physical interest the conservation of spin is restored.

\subsubsection{Mirror Symmetry}

If the junction has a mirror symmetry under interchanging leads $(1,2)
\leftrightarrow (3,4)$ or $(1,4) \leftrightarrow (2,3)$, it follows that
\begin{equation}
G_{XY} = 0.
\end{equation}
Though mirror symmetry is not generically present in a point contact we will
argue that that symmetry is restored in the low energy fixed points of
interest.  Moreover, the {\it crossover} between the critical fixed point and
the stable fixed point described by (\ref{scalingform}) is also along a line with mirror
symmetry.  Thus the crossover conductance is characterized by {\it two}
parameters, $G_{XX}$ and $G_{YY}$, which are simply the two terminal
conductances.

\subsubsection{Critical conductance}

At the transition, where the point contact is just being pinched off the two
terminal conductances must be equal,
\begin{equation}
G_{XX} = G_{YY} \equiv G^*.
\end{equation}
In addition, we will argue that this fixed point also has
spin rotational symmetry and mirror symmetry.
Thus, the critical four terminal conductance $G_{ij}$ depends on a {\it single}
parameter $G^*$.

\subsection{Relation to Kubo conductance}

In this section we relate the conductance matrix $G_{AB}$ to
the conductances of the SLL
model, which can be computed with the Kubo formula.  There are two
issues to be addressed.  First is to translate $G_{AB}$
into the spin and charge conductances of the SLL
model.  Second, we must relate the physical conductance measured with
leads to the conductance computed with the Kubo formula.  The Kubo
conductance describes the response of an {\it infinite} Luttinger
liquid, where the limit $L\rightarrow\infty$ is taken {\it before}
$\omega\rightarrow 0$.  This does not take into account the contact
resistance between the Luttinger liquid and the electron reservoir
where the voltage is defined.  An appropriate model to account for
this is to consider a 1D model for the leads in which the Luttinger
parameter $g=1$ for $x>L$\cite{maslov,safi}.

In this section we assume time reversal symmetry and
that spin is conserved.  In this case we
may define the charge and spin currents in the Fermi liquid leads ($x>L$)
to be,
\begin{eqnarray}
I_\rho &= I_{1,\rm in} + I_{4,\rm in} - I_{1,\rm out}- I_{4,\rm out}
\nonumber\\
I_\sigma &= I_{1,\rm in} - I_{4,\rm in} + I_{1,\rm out} - I_{4,\rm
out}.
\end{eqnarray}
Similarly, define charge and spin voltages
\begin{eqnarray}
V_\rho &= (V_1 + V_4 - V_2 - V_3)/2\nonumber \\
V_\sigma &= (V_1 - V_4 + V_2 - V_3)/2.
\end{eqnarray}
These are related by the conductance matrix.
\begin{equation}
\label{ialphavbeta}
I_\alpha = G_{\alpha\beta} V_\beta,
\end{equation}
where $\alpha,\beta = \rho,\sigma$.  By comparing (\ref{iavb}) and (\ref{ialphavbeta}) it is clear
that
\begin{eqnarray}
\label{gxgrho}
G_{XX} &=& G_{\rho\rho}\nonumber \\
G_{YY} &=& 2 e^2/h - G_{\sigma\sigma} \\
G_{XY} &=& G_{\rho\sigma} = - G_{\sigma\rho}.\nonumber
\end{eqnarray}

$G_{\alpha\beta}$ can be computed using the Kubo formula using the model
in which the interactions are turned off for $x>L$.  It is useful,
however, to relate this to the Kubo conductance $G^K_{\alpha\beta}$ of an infinite
Luttinger liquid.  This can be done by relating the voltage
$V_{\alpha = \rho,\sigma}$ of the Fermi liquid leads with $g_\rho=g_\sigma = 1$
to the voltage $\bar V_\alpha$ of the incoming chiral modes of the
Luttinger liquid with $g_\rho = g$ and $g_\sigma = 1/g$.
By matching the boundary conditions
at $x=L$ this contact resistance has the form
\begin{equation}
\label{tildev v}
\tilde V_\alpha - V_\alpha = R^c_{\alpha\beta} I_\beta
\end{equation}
with
\begin{equation}
\label{rc}
R^c_{\alpha\beta} = {h\over e^2}{{g_\alpha-1}\over{2 g_\alpha}} \delta_{\alpha\beta}.
\end{equation}
The Kubo formula with infinite leads relates $I_\alpha =  G^K_{\alpha\beta}
V_\beta$.  Eliminating $\bar V_\alpha$ from (\ref{tildev v}) and (\ref{rc}) gives the matrix
relation\cite{chamon2}
\begin{equation}
\label{ggkubo}
G_{\alpha\beta} = \left[ \left(I - R_c G^K\right)^{-1}  G^K
\right]_{\alpha\beta}.
\end{equation}

When there is mirror symmetry, so that $G_{XY}=\mu_{\rho\sigma}=0$,
the conductance matrix is diagonal, so that (\ref{ggkubo}) simplifies.  In that
case we find
\begin{eqnarray}
\label{gxxgkubo}
G_{XX} &=& { G^K_{\rho\rho} \over {1- R_{\rho\rho} G^K_{\rho\rho}}} \nonumber\\
G_{YY} &=& 2{e^2\over h} - { G^K_{\sigma\sigma}
\over {1- R_{\sigma\sigma} G^K_{\sigma\sigma}}}.
\end{eqnarray}

\end{appendix}


\begin{thebibliography}{10}
\bibitem{grhonote}
Here our definition of $g_\rho$ and $g_\sigma$ differes by a factor of 2 from
Ref. \onlinecite{kf1}.
\bibitem{varphi}
In Ref. \onlinecite{kf1} $\tilde\theta_{\sigma,\rho}$ was referred to as
$\varphi_{\sigma,\rho}$.  To avoid confusion with the $\varphi_\sigma$ in section
II.B.5 we follow the notation of Ref. \onlinecite{affleck1}.
\bibitem{similar} A similar estimate was given in Ref.
\onlinecite{oreg} without the numerical factor in the log.
\end{thebibliography}
\end{document}